\pdfoutput=1 

\documentclass{new_tlp}
\usepackage{mathptmx}
\usepackage{listings}
\usepackage{todonotes}
\usepackage{multirow}
\usepackage{tikz}
\usepackage{amsmath}
\usepackage{amsfonts}
\usepackage[draft]{fixme}
\usepackage{url,xspace}
\usepackage{booktabs}
\usepackage{wrapfig}
\usepackage[normalem]{ulem}
\usepackage[threshold=0]{csquotes}
\usepackage{ifthen}
\usetikzlibrary{shapes.misc}

\newcommand{\isarxiv}[0]{dummy}      

\usepackage[xcolor]{changebar}
\cbcolor{blue}
\nochangebars

\newcommand\bcmdtab{\noindent\bgroup\tabcolsep=0pt%
  \begin{tabular}{@{}p{10pc}@{}p{20pc}@{}}}
\newcommand\ecmdtab{\end{tabular}\egroup}

\newcommand{\mi}[1]{\ensuremath{\mathit{#1}}}
\newcommand{\code}[1]{\lstinline@#1@}
\newcommand{\totau}[1]{\stackrel{#1}{\to}}

\newcommand{\m}[1]{\mathit{#1}}
\newcommand{\transITS}[0]{\ensuremath{\m{eqs}}}
\newcommand{\transInner}[0]{\ensuremath{\m{eqsC}}}
\newcommand{\entry}[0]{\ensuremath{\m{entry\_loc}}}
\newcommand{\sizean}[0]{\ensuremath{\m{size}}}
\newcommand{\its}[0]{\ensuremath{\m{ts}}}
\newcommand{\mycomment}[1]{}

\newcommand{\lr}[1]{\langle {#1} \rangle}

\newcommand{\secbeg}{\vspace{-0.1cm}}

\newcommand{\toolname}{\textsf{MaxCore}\xspace}
\newcommand{\cofloco}{\textsf{CoFloCo}\xspace}
\newcommand{\coflococ}{\textsf{CoFloCo$_C$}\xspace}
\newcommand{\pubs}{\textsf{PUBS}\xspace}
\newcommand{\pubsc}{\textsf{PUBS$_C$}\xspace}

\newcommand{\verymax}{\textsf{VeryMax}\xspace}
\newcommand{\aprove}{\textsf{AProVE}\xspace}
\newcommand{\maxcorecoflocos}{\textsf{MaxCore(C)}\xspace}
\newcommand{\maxcorepubss}{\textsf{MaxCore(P)}\xspace}

\newcommand{\loopus}{\textsf{Loopus}\xspace}

\newcommand{\sccproc}[0]{\ensuremath{\mi{scc}}}

\newcommand{\constraints}[0]{\ensuremath{\mi{Ct}}}
\newcommand{\locs}[0]{L}
\newcommand{\circled}[1]{\textcircled{\tiny #1}}
\newcommand{\clp}[0]{\ensuremath{CLP(\mathbb{Z})}}
            
\newtheorem{definition}{Definition}
\newtheorem{theorem}{Theorem}
\newtheorem{corollary}{Corollary}
\newtheorem{example}{Example}

  \title[Resource Analysis driven by Termination Proofs] {Resource
        Analysis driven by\\ (Conditional) Termination
        Proofs\footnote{This work was funded partially by the Spanish
        MICINN/FEDER, UE projects RTI2018-094403-B-C31,
        RTI2018-094403-B-C33 and RTI2018-095609-B-I00, the MINECO
        project TIN2015-69175-C4-2-R, the MINECO/FEDER, UE projects
        TIN2015-69175-C4-3-R and TIN2015-66293-R, and by the CM
        project S2018/TCS-4314.}}

  \author[E. Albert, M. Bofill, C. Borralleras,  E.
  Martin-Martin \and
   A. Rubio]
         {ELVIRA ALBERT$^1$, MIQUEL BOFILL$^2$, CRISTINA BORRALLERAS$^3$ \and
           ENRIQUE MARTIN-MARTIN$^1$, ALBERT RUBIO$^1$\\
  $^{1}$ DSIC, Complutense University of Madrid (UCM), E-28040 Madrid, Spain\\
  $^{2}$ IMAE, University of Girona (UdG), E-17003 Girona, Spain\\
  $^{3}$ University of Vic - Central University of Catalonia (UVic-UCC), 08500 Vic (Barcelona), Spain\\
}   

\jdate{May 2019}
\pubyear{2019}
\pagerange{\pageref{firstpage}--\pageref{lastpage}}
\doi{XXX}

\begin{document}

\label{firstpage}

\maketitle

 \begin{abstract}
   When programs feature a complex control flow, existing techniques
   for resource analysis produce \emph{cost relation systems} (CRS)
   whose cost functions retain the complex flow of the program and,
   consequently, might not be solvable into closed-form \emph{upper
     bounds}. This paper presents a novel approach to resource
   analysis that is driven by the result of a termination
   analysis. The fundamental idea is that the termination proof
   encapsulates the flows of the program which are relevant for the
   cost computation so that, by driving the generation of the CRS
   using the termination proof, we produce a
   \emph{linearly-bounded} CRS (LB-CRS). A LB-CRS is composed of cost
   functions that are guaranteed to be \emph{locally} bounded by linear
   ranking functions and thus greatly simplify the process of CRS solving. We
   have built a new resource analysis tool, named \toolname, that is
   guided by the \verymax termination analyzer and uses 
   \cofloco and \pubs as CRS solvers.  Our experimental results on the set of
   benchmarks from the Complexity and Termination Competition 2019 for
   C Integer programs show that \toolname outperforms
 all   other resource analysis tools.
    \ifthenelse{\isundefined{\isarxiv}}{}{\emph{Under consideration for acceptance in TPLP.}}
  \end{abstract}




\secbeg \secbeg 
\secbeg

\section{Motivation and Related Work}\label{sec:introduction}


The classical approach to resource analysis by 
\citeANP{DBLP:journals/cacm/Wegbreit75} consists of two steps: (1)
the generation of a cost relation system (CRS) from the program that
defines by means of recursive cost functions its resource consumption,
(2) solving the CRS into a closed-form expression that bounds its
cost. This approach is generic w.r.t.\ the \emph{cost model} that
defines the type of resource that is being measured, e.g., it has been
applied to estimate number of execution steps, memory, energy
\cite{DBLP:conf/fopara/LiqatGK0GHE15,DBLP:conf/scopes/GrechGPKME15}, user-defined cost models
\cite{DBLP:conf/iclp/NavasMLH07}. W.l.o.g., we use the cost model
adopted in the Complexity and Termination competition
\url{http://termination-portal.org/wiki/Termination_Competition_2019}
(abbreviated as TermComp) which simply estimates the asymptotic
complexity order (e.g., by accumulating constant values in cost functions). This classical resource analysis approach has been
applied to a wide variety of declarative and imperative programming
languages: earlier work applied it to
functional \cite{DBLP:journals/cacm/Wegbreit75} and logic languages
\cite{DBLP:journals/toplas/DebrayL93,DBLP:conf/sas/DebrayGHL94}, later
work to imperative languages such as Java and Java bytecode
\cite{AlbertAGPZ07-short}, concurrent programs
\cite{DBLP:conf/ppdp/GarciaLL15,AlbertCPJR18}, LLVM
\cite{DBLP:conf/scopes/GrechGPKME15,DBLP:conf/fopara/LiqatGK0GHE15},
among others. In most cases, the program written in any
imperative/declarative, source/bytecode language is first transformed
into a simpler intermediate representation (IR) that works only on Integer
data, which is the starting point of our work. For this, a size abstraction is applied on the program to
transform all data into their sizes (e.g., by using the well-known
term-size/term-depth abstractions, or the path-length norm  by
\citeN{DBLP:journals/toplas/SpotoMP10} for
heap-allocated data structures, etc). 
This step is followed by a size analysis
\cite{DBLP:conf/popl/CousotH78} that infers size relations among the
program variables. 
Therefore, step (1) above can be conceptually split into two parts:
(1a) the transformation of the program into an  Integer
IR using a language-specific size
abstraction, 
and (1b) the generation of a CRS from the IR
using the gathered size relations. The IR we adopt in
the paper are Integer Transition Systems (abbreviated as TS)
which 
are an official input language for TermComp.
For the sake of generality, our work assumes that the input program
(written in any language) has been already transformed into a TS and a
language-specific size analysis has been applied, 
and focuses on (1b).

\begin{figure}[t]\secbeg  
\fbox{
~~~\begin{minipage}{5.2cm}
\begin{lstlisting}[language=C]
int aaron3(int x,int y,
           int z,int tx) {
  while (x >= y) {
    if (nondet() > 0) {
      z = z - 1;
      tx = x;
      x = nondet();
      if (x > tx+z) return 0;
    }
    else {
      y = y + 1;
    }
  }
  return 0;
}
\end{lstlisting}	
\end{minipage}
\begin{minipage}{7cm}
~~~~~~~~~~\begin{tikzpicture}[auto,>=latex,scale=.7,transform shape] 
\tikzstyle{location}=[draw,circle]

\node[location] (0) {$l_0$};
\node[location] (1) [right of = 0, node distance = 2cm] {$l_1$};
\node[location] (2) [right of = 1, node distance = 2.4cm] {$l_2$};
\path[->] (0) edge node {$\tau_0:\m{true}$} (1);

\path[->] (1) edge [loop above] node {\hspace{-0.8cm}$
  \begin{aligned}
    \tau_1:\; & x\geq y \\
    & \m{undf1} > 0\\
    & \m{undf2} < x + z\\
    & x' = \m{undf2}\\
    & z' = z - 1
  \end{aligned}
$} (1);

\path[->] (1) edge [loop below] node {\hspace{-0.95cm}$
  \begin{aligned}
    \tau_2:\; & x\geq y \\
    & \m{undf1} \leq 0\\
    & y' = y + 1
  \end{aligned}
$} (1);

\path[->] (1) edge [bend left] node[very near end, above = 2mm of 2] {\hspace{1cm}$
  \begin{aligned}
    \tau_3:\; & x\geq y \\
    & \m{undf1} > 0\\
    & \m{undf2} \geq x + z\\
    & x' = \m{undf2}\\
    & z' = z - 1 
  \end{aligned}
$} (2);

\path[->] (1) edge [bend right] node[swap, below = 1mm] {$\tau_4: x < y$} (2);

\end{tikzpicture}\vspace{-0.4cm}
\[ \vspace{-0.1cm}
\begin{array}{@{~}l@{~}c@{~}l@{~}l@{~}}\hline
c_{l_0}(X,Y,Z,Co) & \textnormal{:-} & \tau_0, c_{l_1}(X,Y,Z,Co'),Co \textnormal{~\#=~}Co'+1. \\ 

c_{l_1}(X,Y,Z,Co) & \textnormal{:-} & \tau_1, c_{l_1}(X',Y,Z',Co'),Co\textnormal{~\#=~}Co'+1.\\
c_{l_1}(X,Y,Z,Co) & \textnormal{:-} & \tau_2, c_{l_1}(X,Y',Z,Co'),Co\textnormal{~\#=~}Co'+1. \\ 
c_{l_1}(X,Y,Z,Co) & \textnormal{:-} & \tau_3, c_{l_2}(X',Y,Z',Co'),Co\textnormal{~\#=~}Co'+1. \\ 
c_{l_1}(X,Y,Z,Co) & \textnormal{:-} & \tau_4, c_{l_2}(X,Y,Z,Co'),Co\textnormal{~\#=~}Co'+1.\\ 
c_{l_2}(X,Y,Z,Co) & \textnormal{:-} & Co\textnormal{~\#=~}1. \\
%
%
%
%
%
\end{array}
\]
\end{minipage}} 
\caption{Motivating example (left). Direct TS (upper-right). Non-solvable CRS
  (bottom-right)} \label{running}\vspace{-0.4cm}

\end{figure}
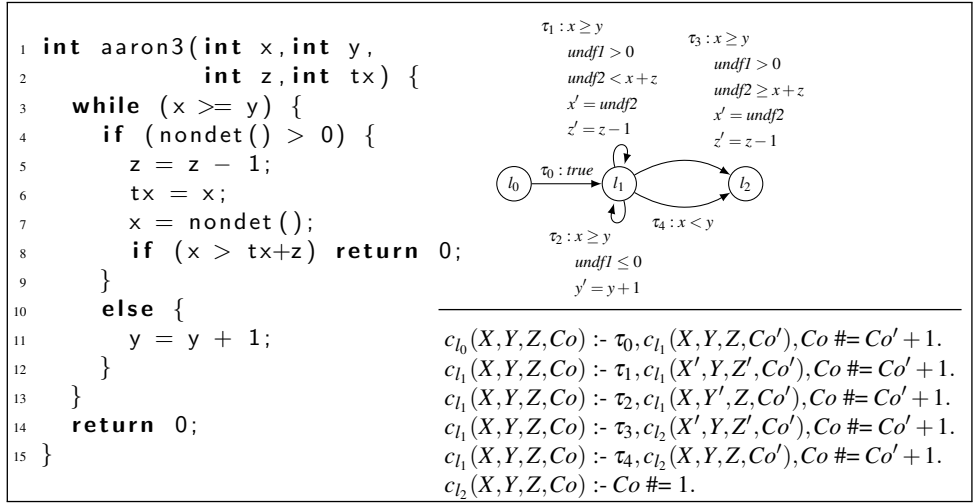

An important limitation of this classical approach to resource
analysis is that CRS inherit the structure of the input program from
which they are generated or, equivalently, of its IR. When the program features a complex control flow, this
might lead to CRS that cannot be solved in step (2). Our motivating
example is \code{aaron3}, borrowed from the set of benchmarks used in
 TermComp'19.  Fig.~\ref{running} shows the C implementation for
this program (left) and the TS directly obtained from it
(up-right). The TS will be explained in further detail later, by now,
we only want to emphasize that it comprises the different paths in the
execution flow and that its arrows are labeled with the
\emph{constraints} that are gathered along each path (\emph{undef}
variables are fresh variables used to represent the unknown result of
function \textsf{nondet}). For instance, the upper arrow from $l_1$
represents the iteration of the loop that executes the \textsf{then}
branch of the first \textsf{if} statement and it accumulates the
constraints gathered from those instructions in $\tau_1$ 
\cbstart 
(the guard contains $\mi{undef} < x + z$ instead of $\mi{undef} \leq x + z$ because the condition in line 8 is evaluated after the 3 assignments, so the variable $z$ refers to the original value minus one).
\cbend 
Note that,
 regardless of the programming language used to implement
 \code{aaron3}, a similar TS would be produced. 
\cbstart The CRS, written as a \clp{} program, that has been obtained by a
standard cost analysis from this TS is shown in the figure
(down-right). 
We can observe that the structure of the cost functions (i.e., the predicates)
corresponds directly to the flow in the original TS, with one cost
function per location in the TS and the constraints guarding the cost
equations (i.e., the clauses). The cost accumulated by each function
is calculated in the last parameter of the predicates. While one
could execute this \clp{} program for concrete input values, our purpose
is to obtain an upper bound for $Co$ that is sound for any possible
execution, i.e., solve the CRS into a closed-form upper bound. 
\cbend 
However, this CRS is not solvable by existing systems (e.g.,
\cofloco, \pubs) due to two reasons: (1) they rely on linear ranking
functions to bound the number of iterations that loops (i.e., the
recursive predicates) perform, while
$c_{l_1}$ requires the lexicographic ranking function $\langle z,
x-y\rangle$, and (2) they cannot find the phases in the execution flow
for the different increase/decrease of variables.
Concretely, the loop presents two phases. In the first phase (when $z > 0$), \cbstart at each iteration either $z$ decreases and $x$ takes an arbitrary value smaller than or equal to $x+z$, or $y$ increases by one. In the worst case $x$ increases, and after every increment of $x$ there may be $x-y$ increments of $y$ followed by a new update of $x$. However, these potential increments in $x$ \cbend can only happen $z$ times, and then the loop enters the second phase where $z \leq 0$. In this other phase, $x$ decreases or $y$ increases, therefore reducing the difference $x-y$ at each iteration. 
 

The problem of the non-solvability of the CRS obtained from complex
flow programs was  observed in
\cite{DBLP:conf/aplas/Flores-MontoyaH14}, which
proposes to partition all possible executions of the program into a
finite set of execution patterns, named \emph{chains}, so that more
precise constraints can be inferred for each of the chains,  \cbstart that \cbend
results in simpler ranking functions and more upper bounds being found. However, the computation of
the chains is not guided by semantic criteria, rather a full
partitioning is carried out, \cbstart that \cbend might lead 
to inaccuracy as our example shows. Indeed,  \cofloco 
\cite{DBLP:phd/dnb/Montoya17} ---implementing the
chains--- is not able to infer an upper bound: it
detects 5 different chains for the loop in \code{aaron3} but can only
infer a bound for 2. One of those detected chains
is the loop formed by the transitions with constraints $\tau_1$ and then $\tau_2$. This chain is detected for the precondition $x\geq y$, which is not strong enough to obtain a linear ranking function.
Since the chain detection was not able to extract the finer phases
above depending on the value of $z$, \cofloco cannot find an upper
bound.  \cbstart Further related work based on finding phases includes
\cite{DBLP:conf/pldi/GulwaniJK09,DBLP:conf/cav/SharmaDDA11}. The former is based on size-change
constraints that are less expressive than the general linear
constraints used by \cite{DBLP:conf/aplas/Flores-MontoyaH14} and
us.  The latter computes rather sophisticated phases but its main
target is on proving safety properties, and it is
unclear how effectively it would perform for cost.
\cbend 

The main idea of our approach that differs from such previous
work is to use, as semantic criterion to guide the CRS
generation, the termination proofs inferred by a powerful termination
analyzer as they comprise the actual phases needed to compute resource
bounds.  This idea is materialized in our analysis by transforming the
TS into a \emph{hierarchically loop-nested} TS that witnesses all
components in the termination proof (e.g., the one for \code{aaron3}
appears later in Fig.~\ref{TSrunning}). The benefit of hierarchically
loop-nested TS is that they allow us to produce CRS that are
\emph{Linearly-Bounded} (LB), as shown later in
Fig.~\ref{fig:finalCRS}. Cost functions in the LB-CRS are guaranteed
to have linear ranking functions. Thus, the
solving process is greatly
simplified, e.g., we indeed find an upper
bound \cbstart 
of $O(n^3)$ 
for \code{aaron3}, where $n$ is the maximum of the parameters
\code{x}, \code{y}, \code{z}, and \code{tx}. \cbend Interestingly, we
rely on a \emph{conditional} termination analysis
\cite{DBLP:conf/tacas/BorrallerasBLOR17} that, when it cannot prove
termination unconditionally, tries to infer preconditions under which
termination is guaranteed. Conditional termination proofs allow us to
generalize our results to conditional upper bounds.  \cbstart Finally,
another work related to ours is 
\cite{SZV14:CAV}.  The  similarity with our approach is that both
can use lexicographic ranking functions to bound the cost but our technique is more
general as it allows more powerful termination arguments, besides 
not being limited to difference constraints as 
\cite{SZV14:CAV}.   According to our experimental results, the precision
of our system significantly outperforms their system \loopus.


\cbend



\emph{Summary of contributions.} 
 Briefly, the main
 contributions of our work are:
(i) We define the concept of \emph{lexicographic phase-level
  termination proof}, $\m{Proof}$, to store information on the phases
which have been considered during the conditional termination proof 
and unfold the TS
accordingly.  (ii) We present a
transformation which takes the unfolded TS together with the
$\m{Proof}$s of its phases
and produces a hierarchically loop-nested TS$^h$ which explicitly
represents the different components of the termination proof.
 The CRS generated from TS$^h$ is
  \emph{locally} LB, 
  although still needs to be globally bounded in the solving step.
  (iii) We propose extensions of the basic framework: to embed the
  ranking functions into the CRS; and to embed the preconditions
  inferred by the termination analysis so that
conditional upper bounds can be generated.
(iv)
We implement \toolname (standing for Max-SMT based termination
analyzer + COst Recurrence Equation solver), that makes use of
\verymax \cite{DBLP:conf/tacas/BorrallerasBLOR17} to generate the
conditional termination proofs from which our implementation produces
CRS, and uses both \cofloco and \pubs as backend solvers.  (v)
We prove experimentally on the benchmarks from TermComp'19 for C
Integer programs that \toolname outperforms all existing resource
analyzers in number of: problems solved, unique problems solved, more
accurate solutions, and overall score.


\secbeg 
\secbeg 
 
\section{Lexicographic Phase-Level Termination Proofs and Unfolded TS}\label{sec:lexic-phase-level}

In this section we present an overview
of~\cite{DBLP:conf/tacas/BorrallerasBLOR17} and propose how to adapt
the results of this analysis to guide the generation of the
CRS. Essentially, \citeN{DBLP:conf/tacas/BorrallerasBLOR17}
describe a template-based method for proving conditional
termination, and then  show how to use conditional proofs to
advance towards an (unconditional) termination proof. The key idea is
that conditional termination proofs show termination for a subset of
states which can be excluded in the rest of the termination analysis,
i.e., the rest of the proof can concentrate on the complementary
states. This way, the method allows generating not only a termination
proof, but also a characterization of the \emph{execution phases} in a
program. An execution \emph{phase} characterizes a subset of states in which
termination follows from a different conditional invariant.

We assume programs are given as  \emph{(Linear) Integer Transition
  System}s (TSs). 
A TS is a control-flow graph with transitions $\tau$ of the form
$(l_s,\rho,l_t)$, where $l_s$ and $l_t$ are locations and $\rho$ is a
conjunction of \emph{linear} inequalities describing the transition
relation (by abuse of notation we sometimes use $\tau$ to express only
its associated $\rho$). When the input program contains non-linear
instructions that are not handled by our analysis, they are translated
into \emph{undefined} variables within the inequalities to express the
loss of information.  \cbstart For instance, if the condition in the
\texttt{if} statement in line 4 was \texttt{x*y}, this is transformed
into a call to function \texttt{nondet} that  has led to the
introduction of the undefined variable \emph{undef1} (representing
the unknown value 
\texttt{x*y}) in the constraints $\tau_1$, $\tau_2$ and $\tau_3$.
\cbend The formula $\rho$ can contain primed variables $v'$, which
represent the value of a variable $v$ after the transition (equalities
$v'=v$ are omitted).  A \emph{program component} $C$ of a program $P$
is the set of transitions of a \emph{strongly connected component}
(SCC) of the CFG of $P$. For example, in the TS of Fig.~\ref{running}
there are two trivial (i.e., single node) SCCs; the transitions
$\tau_1$ and $\tau_2$ form a non-trivial program component.

Termination of a program is proven component-by-component, and
termination of a program component is proven iteratively by removing
transitions that can only be finitely executed. A \emph{ranking
  function} for a component $C$ and a transition
$\tau=(l_s,\rho,l_t)\in C$ is a function
$R: \mathbb{Z}^n\to \mathbb{Z}$ 
such that it is bounded from below $\rho\models R\geq 0$, it
strictly decreases $\rho\models R > R'$ and, for every
$(\hat{l}_s,\hat{\rho},\hat{l}_t)\in C$, it is non-increasing
$\hat{\rho}\models R\geq R'$, where $R'$ is the version of $R$ using
primed variables. The key property of ranking functions is that if one
transition admits one, then it cannot be executed infinitely.
In our setting, proving termination of a component $C$ is based on finding a linear
ranking function together with some supporting invariants that ensure
the conditions for being a ranking function. Invariants
are described by a function $Q: {\cal L}(C)\to {\cal F}(V)$, where
${\cal L}(C)$ is the set of locations of $C$ and ${\cal F}(V)$ are 
conjunctions of linear inequalities over the variables $V$ of the
program.  Then, strictly decreasing transitions w.r.t. this ranking
function can be removed and the process is iterated over the remaining
SCCs. However, although all supporting
invariants are inductive in~\cite{DBLP:conf/tacas/BorrallerasBLOR17}, they are not necessarily initiated in all
computations. In this case, those
invariants are called \emph{conditional invariants} as they yield a
precondition for termination, i.e., they prove termination for a
subset of initial states. Therefore, the rest of the proof can be restricted to the remaining states. This makes the proof method more
powerful and, as a by-product, loops with different execution phases
can be handled naturally.

In this paper, we propose to store information on the
phases which have been considered during the termination proof,
together with the lexicographic termination proof of each phase. This
information will capture all the possible execution flows in the
execution of the program and will be used for guiding the generation
of the CRS.

\begin{definition}[lexicographic phase-level termination proof, $\m{Proof}$]
  \label{def-lexproof}
Let $C$ be a component and $R$ a ranking function for $C$ with a
supporting conditional invariant $Q$. Then $C$ can be split into
$C^> \uplus C^\m{subSCC} \uplus C^\m{noSCC}$ where:
\secbeg \begin{itemize}
\item $C^>$ contains the strictly-decreasing transitions in $C$
  w.r.t. $R$ assuming $Q$, 
\item $C^\m{subSCC}$ contains the transitions that belong to an $SCC$
  in $C\setminus C^>$, and 
\item $C^\m{noSCC}=C\setminus (C^>\uplus C^\m{subSCC})$ contains the
  transitions that after removing the strictly decreasing transitions
  do not belong to any $SCC$. 
\end{itemize}\secbeg\secbeg
We denote by $C^R$ the set of transitions $C^>\uplus C^\m{noSCC}$.
A lexicographic phase-level termination proof for (a phase of) $C$ can
be represented by a tree-like structure
$\m{Proof}(C)=\langle R,Q,C^R,$ $\langle \m{Proof}$ $(C_1),$
$\ldots,\m{Proof}(C_k)\rangle\rangle$ where $C_1,\ldots,C_k$ are the
new SCCs in $C^\m{subSCC}$.
\end{definition}
The information kept for the termination proof of (a phase of) a
component in the definition above is (i) the ranking function used;
(ii) its supporting conditional invariants; (iii) the set of
transitions removed, either because they strictly decrease
wrt. the ranking function or they do not belong to any SCC after
removing the strictly decreasing ones; and, recursively, (iv) the
information corresponding to the termination proof of the remaining
SCCs after transition removal.

\begin{example}
  \label{runningproof}
  Let us consider the non-trivially terminating component
  $C=\{\tau_1,\tau_2\}$ of Fig.~\ref{running}, where
  $\tau_1 = (l_1,\rho_1,l_1)$, $\tau_2 = (l_1,\rho_2,l_1)$,
  $\rho_1 = x\geq y \land \m{undf1} > 0 \land \m{undf2} < x + z \land
  x' = \m{undf2} \land z' = z - 1$ and
  $\rho_2 = x\geq y \land \m{undf1} \leq 0 \land y' = y + 1$. In this
  case a possible ranking function is $x-y$, with supporting
  conditional invariant $z\leq 0$. In particular, we have $x-y\geq 0$
  both in $\tau_1$ and $\tau_2$, and $x-y$ strictly decreases in
  $\tau_2$, as well as in $\tau_1$ assuming $z\leq 0$. Therefore we
  have $C^> = \{\tau_1,\tau_2\}$, $C^\m{subSCC} = \emptyset$ and
  $C^\m{noSCC} = \emptyset$, giving us $C^R = \{\tau_1,\tau_2\}$ and
  $\m{Proof}(C) = \langle x-y, Q, \{\tau_1,\tau_2\}, \langle\rangle
  \rangle$ where $Q(l_1) = z \leq 0$.
  This is a conditional termination proof for $C$ with supporting
  conditional invariant $z\leq 0$.  To complete the termination proof,
  we have to analyze the rest of the states where $z\geq 1$. For this,
  we will assume an entry transition of the form
  $(l_0, z\geq 1, l_1$) instead of the original
  $(l_0, \m{true}, l_1)$, and a strengthened version of $C$ defined by
  $C'=\{\tau_1',\tau_2'\}$, with
  $\tau_1' = (l_1,\rho_1',l_1)$,
  $\tau_2' = (l_1,\rho_2',l_1)$,
  $\rho_1' = \rho_1\land z\geq 1$ and
  $\rho_2' = \rho_2\land z\geq 1$.
  In this new phase, $z-1$ is a ranking function for $C'$ and
  $\tau_1'$ without the need of any additional supporting invariant,
  since $z-1\geq 0$ in $\tau_1'$, $z-1$ strictly
  decreases in $\tau_1'$ and it is non-increasing in
  $\tau_2'$. \cbstart Therefore, we have $C^{'>} = \{\tau_1'\}$, $C^\m{'subSCC} = \{\tau_2'\}$ and
  $C^\m{'noSCC} = \emptyset$ and \cbend
  $\m{Proof}(C') = \langle z-1, Q', \{\tau_1'\}, \m{Proof}(\{\tau_2'\})
  \rangle$, with $Q'(l_1) = \m{true}$.
  Finally, $x-y$ is a ranking function for $\tau_2'$, giving
  $\m{Proof}(\{\tau_2'\}) = \langle x-y, Q', \{\tau_2'\},
  \langle\rangle \rangle$.
\end{example}

\begin{wrapfigure}[12]{l}{0.2\textwidth} \vspace{-0.4cm}
  \begin{minipage}{.2\textwidth}
    \centering 
    \begin{tikzpicture}[auto,>=latex,scale=.7,transform shape]
\tikzstyle{location}=[draw,circle,minimum size=7mm]

\node[location] (0) {$l_0$};
\node[location] (1) [above right = 1.7cm and 1.2cm of 0] {$l_1$};
\node[location] (2) [below right = 1.7cm and 1.2cm of 1] {$l_2$};
\node[location] (11) [below right = 1.7cm and 1.2cm of 0] {$\widehat{l_1}$};

\path[->] (0) edge node [sloped, very near end] {$\tau_0\land z'\geq 1$} (1);
\path[->] (0) edge node [swap, sloped, very near end] {$\tau_0\land z'\leq 0$} (11);
\path[->] (1) edge [bend left=15] node [sloped, near start] {$\tau_3$} (2);
\path[->] (1) edge [bend right=15] node [sloped, near start] {$\tau_4$} (2);
\path[->] (11) edge [bend left=15] node [swap, sloped, near start] {$\tau_3$} (2);
\path[->] (11) edge [bend right=15] node [swap, sloped, near start] {$\tau_4$} (2);
\path[->] (1) edge [loop above, out=125, in=170, looseness=6] node
[xshift=-2mm, yshift=1mm] {$\tau_1\land z\geq 1$} (1);
\path[->] (1) edge [loop above, out=55, in=10, looseness=6] node
[xshift=2mm, yshift=1mm] {$\tau_2\land z\geq 1$} (1);
\path[->] (11) edge [loop below, out=235, in=190, looseness=6] node
[xshift=-2mm, yshift=-1mm] {$\tau_1\land z\leq 0$} (11);
\path[->] (11) edge [loop below, out=305, in=350, looseness=6] node
[xshift=2mm, yshift=-1mm] {$\tau_2\land z\leq 0$} (11);
\path[->] (1) edge node [rotate=-90, xshift=-5mm, yshift=3mm] {$z\leq 0$} (11);

\end{tikzpicture}
  \end{minipage} \vspace{-0.7cm} \caption{Unfolded TS}  \label{fig:unfolding1}
\end{wrapfigure}
\noindent Lexicographic phase-level termination proofs can be considered to use
a semantically equivalent unfolded version of the TS.  The unfolding
goes as follows. For each transition $(l_s,\rho,l_t)$ of a component
$C$, on the one hand $(l_s,\rho,l_t)$ is strengthened with the
negation of the conditional invariant $Q$ for $C$; more precisely, the
transition is replaced by $(l_s,\rho\land\neg Q(l_s),l_t)$, or a set of
transitions if $\neg Q(l_s)$ has disjunctions. On the other hand, a
transition $(\widehat{l_s},\rho\land Q(l_s),\widehat{l_t})$ is added
between two fresh locations $\widehat{l_s}$ and $\widehat{l_t}$.
Transitions strengthened with the negation of the conditional
invariant correspond to a phase for which termination has not yet been
proven, whereas transitions strengthened with the conditional
invariant correspond to a phase for which termination has already been
proven.  Under this assumption, the remaining proof can be restricted
to transitions strengthened with the negated invariant. A single
transition $(l_s, Q(l_s),\widehat{l_s})$ is added  to
connect the two phases, i.e., to allow switching to a phase for which
termination has already been proven.
Finally, to preserve semantic equivalence of the unfolded transition
system, the entry transitions $(l_s,\rho,l_t)$ of $C$ are unfolded
into $(l_s,\rho\land\neg Q(l_t)',l_t)$ and
$(l_s,\rho\land Q(l_t)',\widehat{l_t})$, while exit transitions are
unfolded into $(l_s,\rho,l_t)$ and $(\widehat{l_s},\rho,\l_t)$.
It is worth noticing that this unfolding is equivalent to the one
described in~\cite{DBLP:conf/tacas/BorrallerasBLOR17} but in general
leads to a simpler TS.
In what follows, we assume the original TS has been unfolded as
described above, and denote it $\m{TS}_{u}$.
Fig.~\ref{fig:unfolding1} shows the unfolded TS corresponding to the
termination proof of the program in Fig.~\ref{running} with:
$\tau_0:\m{true}$; $\tau_1: x\geq y$, $\m{undf1} > 0$,
$\m{undf2} < x + z$, $x' = \m{undf2}$, $z' = z - 1$;
$\tau_2: x\geq y$, $\m{undf1} \leq 0$, $y' = y + 1$;
$\tau_3: x\geq y$, $\m{undf1} > 0$, $\m{undf2} \geq x + z$,
$x' = \m{undf2}$, $z' = z - 1$; $\tau_4: x < y$. Note that we have
also strengthened the transitions looping in $\widehat{l_1}$ with its
conditional invariant $z\leq 0$.
This graph visualizes the loop phases in the program, which have
been described in Sec.~\ref{sec:introduction}.


\secbeg\secbeg\secbeg

\section{Linearly-Bounded Hierarchically-Loop-Nested Integer
  Transition Systems}\label{sec:transf-ts-with}


\begin{figure}[t]\secbeg
  \centering\secbeg
\fbox{
\begin{minipage}{3.9cm}
  \begin{tikzpicture}[auto,>=latex,scale=.6,transform shape]
\tikzstyle{location}=[draw,circle]
\node[location] (0) {$l_0$};
\node (c0) [left of=0,node distance=1.6cm] {$C_0$};
\node (entry) [above left of=0,node distance=1.5cm] {};
\node[location] (1) [below of=0,node distance=3cm] {$l_1$};
\node[location] (2) [right of=0,node distance=3cm] {$l_2$};
\node (c1) [right of=2,node distance=1.55cm] {$C_1$};
\path[->] (entry) edge (0);
\path[->] (0) edge [bend left=10] node {$\tau_1$} (2);
\path[->] (0) edge [bend left=10,draw=none] node[strike out,draw,xshift=2mm,yshift=-1.25mm] {} (2); 
\path[->] (2) edge [bend left=10] node {$\tau_2$} (0);
\path[->] (2) edge [bend left=10,draw=none] node[strike out,draw,xshift=2.5mm,yshift=1.25mm] {} (0); 
\path[->] (2) edge [loop right] node {$\tau_3$} (2);
\path[->] (0) edge [bend left=10] node {$\tau_4$} (1);
\path[->] (1) edge [bend left=10] node {$\tau_5$} (0);
\path[->] (1) edge [loop left=10] node {$\tau_6$} (1);
\path[->] (2) edge [bend left] node {$\tau_7$} (1);
\path[->] (2) edge [bend left,draw=none] node[rotate=30,strike out,draw,xshift=-2mm,yshift=0.75mm] {} (1); 
\draw[dashed] ([xshift=3pt]2) ellipse (1.1cm and 1cm);
\draw[dashed] (-0.2,-1.5) ellipse (1.3cm and 2.5cm);
\end{tikzpicture}
  \vspace{-0.6cm}
\caption{First $\m{split}$ without~\emph{\ref{sametargettoC0}}}
\label{TSplit1}
 \end{minipage}
}
~~\fbox{
\begin{minipage}{3.9cm}
\begin{tikzpicture}[auto,>=latex,scale=.6,transform shape]
\tikzstyle{location}=[draw,circle]
\node[location] (0) {$l_0$};
\node (c0) [left of=0,node distance=1.4cm] {$C_0$};
\node (entry) [above left of=0,node distance=1.5cm] {};
\node[location] (1) [below of=0,node distance=3cm] {$l_1$};
\node[location] (2) [right of=0,node distance=3cm] {$l_2$};
\path[->] (entry) edge (0);
\path[->] (0) edge [bend left=10] node {$\tau_1$} (2);
\path[->] (2) edge [bend left=10] node {$\tau_2$} (0);
\path[->] (2) edge [loop right] node {$\tau_3$} (2);
\path[->] (0) edge [bend left=10] node {$\tau_4$} (1);
\path[->] (1) edge [bend left=10] node {$\tau_5$} (0);
\path[->] (1) edge [loop left] node {$\tau_6$} (1);
\path[->] (2) edge [bend left] node {$\tau_7$} (1);
\node (aux1) [left of=0] {};
\node (aux2) [above of=0] {};
\node (aux3) [above of=2] {};
\node (aux4) [right of=2,node distance=1.5cm] {};
\node (aux5) [below right of=2,node distance=1.3cm] {};
\node (aux6) [below right of=1] {};
\node (aux7) [below of=1] {};
\node (aux8) [left of=1,node distance=1.5cm] {};
\draw [dashed] plot [smooth cycle] coordinates {(aux1) (aux2)
  (aux3) (aux4) (aux5) ([yshift=-5pt]2.south west)
  ([xshift=2mm]1.north east) (aux6) (aux7) (aux8)};
\end{tikzpicture}
\vspace{-0.6cm}
\caption{Final $\m{split}$}
\label{TSplit2}
\end{minipage}
}
~~\fbox{
\begin{minipage}{3.9cm}
\begin{tikzpicture}[auto,>=latex,scale=.6,transform shape]
\tikzstyle{location}=[draw,circle]
\node[location] (0) {$l_0$};
\node (c0) [left of=0,node distance=1.7cm] {$C_0'$};
\node (entry) [above left of=0,node distance=1.5cm] {};
\node[location] (1) [below of=0,node distance=3cm] {$l_1$};
\node[location] (2) [right of=0,node distance=3cm] {$l_2$};
\path[->] (entry) edge node[xshift=1mm,yshift=-2.5mm] {$n_1'=1$} (0);
\path[->] (0) edge [bend left=10] node[yshift=-2pt] {$\tau_1\land n_1 > 0$} (2);
\path[->] (2) edge [bend left=10] node[yshift=2pt] {$\tau_2\land n_1 > 0$} (0);
\path[->] (2) edge [bend left=40] node[yshift=2pt] {$\tau_7'\land n_1' = 0$} (0);
\path[->] (2) edge [loop right] node[xshift=-7.7mm,yshift=-4.5mm] {$\tau_3\land n_1 > 0$}
(2);
\path[->] (0) edge [bend left=10] node[xshift=-1pt] {$\tau_4\land n_1 > 0$} (1);
\path[->] (0) edge [bend left=40] node[near end,xshift=0.5pt,yshift=3mm] {$\tau_7'\land n_1' = 0$} (1);
\path[->] (1) edge [bend left=10] node[xshift=1pt] {$\tau_5\land n_1 > 0$} (0);
\path[->] (1) edge [loop left] node[xshift=5mm,yshift=3.5mm]
{$\tau_6\land n_1 > 0$}
(1);
\path[->] (1) edge [loop right] node {$\tau_7\land n_1=0\land n_1'=1$} (1);
\node (aux2) [above of=0] {};
\node (aux3) [above of=2] {};
\node (aux4) [right of=2,node distance=1.5cm] {};
\node (aux5) [below of=2,node distance=1.1cm] {};
\node (aux6) [below right of=1] {};
\node (aux7) [below of=1] {};
\node (aux8) [left of=1,node distance=1.8cm] {};
\node (aux9) [below right of=0,node distance=3.4cm] {};
\draw [dashed] plot [smooth cycle] coordinates {(aux2)
  (aux3) (aux4) ([xshift=1.25cm]aux5) (aux9)
  ([xshift=1.4mm]1.north east) (aux6) (aux7) (aux8) ([xshift=1pt]c0.east)};
\end{tikzpicture}
\vspace{-0.6cm}
\caption{Move source of $\tau_7$}
\label{TSMove}
\end{minipage}
}

 \vspace{-0.25cm}
\end{figure}

\begin{figure}[t]\secbeg
  \centering\secbeg
  \begin{tikzpicture}[auto,>=latex,scale=.6,transform shape]
\tikzstyle{location}=[draw,circle]
\node[location] (0') {$l_0'$};
\node (c0') [left of=0',node distance=1.9cm] {$C_0''$};
\node (entry') [above left of=0',node distance=1.7cm] {};
\node[location] (1') [below of=0',node distance=4cm] {$l_1'$};
\node[location] (2') [right of=0',node distance=5.5cm] {$l_2'$};
\node[location] (f0) [right of=1',node distance=4cm] {$f_0$};
\path[->] (entry') edge node[xshift=-0.5mm,yshift=-0.5mm] {$n_1'=1\land n_2'=1$} (0');
\path[->] (0') edge [bend left=10] node[yshift=-1pt] {$\tau_1\land n_1 > 0\land n_2 > 0$} (2');
\path[->] (2') edge [bend left=10] node[yshift=1pt] {$\tau_2\land n_1 > 0\land n_2 > 0$} (0');
\path[->] (2') edge [bend left=35] node[xshift=3mm,yshift=1pt] {$\tau_7'\land n_1' = 0\land n_2 > 0$} (0');
\path[->] (2') edge [loop right] node[xshift=-7.7mm,yshift=-7mm]
{$\begin{aligned}
    &\tau_3\land n_1 > 0\\
    &\land\;n_2 > 0
    \end{aligned}$}
(2');
\path[->] (0') edge [bend left=10] node[xshift=-1pt,yshift=-10pt]
{$\begin{aligned}
    &\tau_4\land n_1 > 0\\
    &\land\;n_2 > 0
  \end{aligned}$}
(1');
\path[->] (0') edge [bend left=40] node[near
end,xshift=-3pt,yshift=0mm]
{$\begin{aligned}
    &\tau_7'\land n_1' = 0\\
    &\land\;n_2 > 0
  \end{aligned}$}
 (1');
\path[->] (1') edge [bend left=40] node[near
start,xshift=3pt,yshift=0mm]
 {$\tau_7'\land n_1 = 0\land n_2' = 0$}
 (0');
\path[->] (1') edge [bend left=10] node[xshift=1pt,yshift=-10pt]
{$\begin{aligned}
    &\tau_5\land n_1 > 0\\
    &\land\;n_2 > 0
    \end{aligned}$} (0');
\path[->] (1') edge [loop left] node[xshift=0mm,yshift=0mm]
 {$\begin{aligned}
     &\tau_6 \land n_1 >0\\
     &\land\;n_2 > 0
   \end{aligned}$}
(1');
\path[->] (0') edge node[near end,xshift=-6mm,yshift=3mm] {$\tau_7'\land n_1 = 0\land
  n_2 = 0$} (f0);
\node (aux2') [above of=0'] {};
\node (aux3') [above of=2'] {};
\node (aux4') [right of=2',node distance=1.6cm] {};
\node (aux5') [below of=2',node distance=1.2cm] {};
\node (aux7') [below of=1'] {};
\node (aux8') [left of=1',node distance=3.5cm] {};
\node (aux9') [below right of=0',node distance=3cm] {};
\draw [dashed] plot [smooth cycle] coordinates {([yshift=3mm]aux2')
  (aux3') (aux4') ([xshift=1.25cm]aux5') (aux9')
  ([xshift=2.3cm]1'.east) ([yshift=4mm]aux7') ([yshift=-4mm]aux8')
  ([yshift=9mm]aux8') ([yshift=-15mm]c0'.south) ([xshift=-10mm,yshift=-2mm]aux2')};
\node[location] (0) [right of=2,node distance=7cm] {$l_0$};
\node (c0) [left of=0,node distance=1.7cm] {$C_0'$};
\node[location] (1) [below of=0,node distance=3cm] {$l_1$};
\node[location] (2) [right of=0,node distance=3cm] {$l_2$};
\path[->] (0) edge [bend left=10] node[yshift=-2pt] {$\tau_1\land n_1 > 0$} (2);
\path[->] (2) edge [bend left=10] node[yshift=2pt] {$\tau_2\land n_1 > 0$} (0);
\path[->] (2) edge [bend left=40] node[yshift=2pt] {$\tau_7'\land n_1' = 0$} (0);
\path[->] (2) edge [loop right] node[xshift=-7.7mm,yshift=-4.5mm] {$\tau_3\land n_1 > 0$}
(2);
\path[->] (0) edge [bend left=10] node[xshift=-1pt] {$\tau_4\land n_1 > 0$} (1);
\path[->] (0) edge [bend left=40] node[near end,xshift=0.5pt,yshift=3mm] {$\tau_7'\land n_1' = 0$} (1);
\path[->] (1) edge [bend left=10] node[xshift=1pt] {$\tau_5\land n_1 > 0$} (0);
\path[->] (1) edge [loop left,out=150,in=120,looseness=8] node[xshift=2mm,yshift=2.5mm]
{$\tau_6\land n_1 > 0$}
(1);
\path[->] (f0) edge [bend left=10] node[xshift=7mm] {$\tau_7\land n_1 = 0\land n_1'=1$} (1);
\path[->] (1) edge [bend left=10] node[yshift=1mm] {$\tau_7'\land n_1=0$} (f0);
\node (aux4) [right of=2,node distance=1.5cm] {};
\node (aux5) [below of=2,node distance=1.1cm] {};
\node (aux7) [below of=1] {};
\node (aux9) [below right of=0,node distance=3.4cm] {};
\draw [dashed] plot [smooth cycle] coordinates {([yshift=2mm]0.north)
  ([yshift=3mm]2.north) (aux4) ([xshift=1.25cm]aux5) (aux9)
  ([yshift=-2mm]1.south east) ([yshift=-2mm]1.south) ([yshift=-2mm]1.south west) (1.west) ([xshift=-3mm,yshift=-1mm]1.north west) ([xshift=-1.8cm,yshift=2mm]1.north west) ([yshift=-8mm]c0.south) ([xshift=5mm]c0.east)};
\end{tikzpicture}
\vspace{-0.6cm}
\caption{
Example of complex transformation} \label{TScomplex} \vspace{-0.3cm}
\end{figure}

The goal of this section is to soundly transform each phase of an unfolded
transition system $\m{TS}_{u}$, which is given as an SCC $C$ with its corresponding $\m{Proof}(C)$ using linear ranking functions
(see Sec.~\ref{sec:lexic-phase-level}) into a TS composed of linearly-bounded hierarchically loop-nested SCCs as
defined below. Let us introduce some notation. By $entryT(C)$,
we denote all entry transitions to $C$, i.e. with target location in
$C$ and source location out of $C$, and by $exitT(C)$ we denote all exit
transitions from $C$, i.e. with source location in $C$ and target
location out of $C$. A location $l$ in $C$ is said to be an
\emph{entry location} if there is a transition in $entryT(C)$ with $l$
as target.  A location $l$ in $C$ is said to be an \emph{exit
  location} if there is a transition in $exitT(C)$ with $l$ as source. In what follows we assume that when we are given a component $C$ we also have $entryT(C)$
and $exitT(C)$.

\begin{definition}[linearly-bounded hierarchically-loop-nested SCC/TS]
  An SCC $C$ is said to be \emph{hierarchically loop-nested} if (i)
  it has a single entry and exit location $e$; (ii) there is a set of
  locations $l_0,\ldots,l_n$ with $e=l_0$ s.t. if $l_i$ is
  connected (with one or more transitions) to another $l_j$ then $j>i$
  or $j=0$ and (iii) for all $l_i$ with $i\geq 0$, either $l_i$ has no
  more connections than these or it is the entry location of a sub-SCC that is
  also a hierarchically loop-nested TS. A TS is
  hierarchically loop-nested if all its subSCCs also are. In addition,
  it is said to be \emph{linearly bounded} if  the loop (with all transitions between
  locations in) $l_0,\ldots,l_n$ is bounded by a linear ranking
  function and all sub-SCCs are
  linearly bounded.
\end{definition}
Therefore, from $C$ and $\m{Proof}(C)$, we aim at generating a
transformed TS that has a representation of nested loops, where every
loop has a single location that is both the entry \cbstart and \cbend the exit
location.  W.l.o.g., we assume that the component $C$ has a single
entry location (if there are several we simply clone $C$ for every
entry location, by renaming locations). Every cloned component $C_i$
for the entry location $i$ will have as entries those of $C$ that have
$i$ as target.  Regarding exit transitions, it is easy to transform any component $C$ with exit locations that are different from the entry 
location into one TS that has only exits from the entry location. \cbstart 
Furthermore, this transformation can be done introducing only transitions from $l_i$ to $l_j$ if a transition from $l_i$ to $l_j$ already exists. This transformation, that we call in what follows $exitToentry$ can be, in general, done to change the source location of a set of transitions from one location to another (and in particular from one exit to an entry). This more general construction, that we call $moveSourceLocation$, takes a component $C$ (including entries), a set of transitions $T$ with the same source location $l$ and a location $e$, and introduces a fresh variable to encode the move from $l$ to the new location $e$ when the transitions in $T$ can be applied, and then changes $T$ to have $e$ as source. The transition system in Fig.~\ref{TSMove} is the result of applying $moveSourceLocation$ to $C_0$, $T=\{\tau_7\}$ and $e=l_1$ in Fig.~\ref{TSplit2}.
\cbend

 Now, we describe how to transform any SCC $C$ and $\m{Proof}(C)$ into
 a LB hierarchically-loop-nested one. As it is a general
 transformation procedure for any possible component $C$, its formal
 description is quite involved. However, in practice, in most cases
 the transformation is not that complex, as we show later in
 Ex.~\ref{runningtransLex} for our running example. We will also provide some examples of the application of the more involved steps.
We first define the following auxiliary function
$\m{split}$ on $C$ which roughly uses $\m{Proof}(C)$ to extract a set
of sub-SCCs (maybe including a single location without transitions)
which represent the inner loops and a subset of $C^R$ (the removed
transitions in the first step of $\m{Proof}(C)$) that are the
transitions performed to go from one inner loop to another and that
form the outer loop. It is important to note that if we remove any of
these selected transitions, the only SCCs of the remaining graph are
the ones we have extracted. The splitting has a DAG-like shape of
components whose \cbstart leaves \cbend return to the unique initial component $C_0$,
and $C_0$ has the same target location for all returning transitions.


\begin{definition}
  Let $C$ be a terminating SCC with $\m{Proof}(C)$. Procedure $\m{split}(C)$
  extracts subcomponents $C_0,\ldots,C_n$ and
  disjoint non-empty sets of transitions $T_0,\ldots, T_n$ with $n\geq 0$, such
  that \renewcommand{\theenumi}{\alph{enumi}}
  \renewcommand{\labelenumi}{\theenumi.}
\secbeg \secbeg \begin{enumerate}
  \item \label{union} the transitions in $C_0,\ldots,C_n$ union $T_0,\ldots, T_n$ coincide with $C$; \cbstart when $C_i$ has no transition, we say it includes the single source location of all transitions in $T_i$,\cbend
    \item \label{union2} every $C_i$ is included in $C'_{j_1}\cup\ldots\cup
      C'_{j_m}\cup C^R$ for some $m\geq 0$, where $\m{Proof}(C'_{j_k})$
      is a subproof of $\m{Proof}(C)$ for $k\in\{1\dots m\}$,
    \item \label{union3} every $T_i$ is included in $C^R$,
    \item \label{split-init} $C_0$ includes the entry location,
    \item \label{split-SCC} every $C_i$ is an SCC and no location is shared between $C_i$ components,
    \item \label{sourceinCi} the source location of all transitions in $T_i$ for $i\in\{0\ldots n\}$ belongs to $C_i$,
    \item \label{dag} the target location of every transition in $T_i$ belongs to some $C_j$ with $j>i$ or to $C_0$,

    \item \label{sametargettoC0} all transitions in $T_0\cup\ldots\cup T_n$ having target location in $C_0$ have the same target location.
    \end{enumerate}\secbeg
\end{definition}
\secbeg\secbeg
As a simple example, the $split$ of the phase where $z\geq 1$ in
our running example (see Fig.~\ref{running}) has one sub-SCC $C_0=\{\tau_2\land z\geq 1\}$ and $T_0=\{\tau_1\land z\geq 1\}$. On the other hand, the $split$ of the phase where $z\leq 0$ has $C_0$ as the trivial SCC containing~$\widehat{l_1}$ and $T_0=\{\tau_1\land z\leq 0,\tau_2\land z\leq 0 \}$.

  Given a terminating SCC $C$ with $\m{Proof}(C)$, the result of
  $\m{split}(C)$ can always be built. 
%
\cbstart
As a possible way to obtain $\m{split}(C)$, we can make a first
selection of $C_0,\ldots,C_n$ and $T_0,\ldots, T_n$ as follows: (i) we
take a set of transitions in $C^R$ having the same source and target
location and remove them from the SCC; (ii) then we recompute the
(maximal) SCCs of the remaining graph, obtaining $C_0,\ldots,C_n$;
(iii) transitions that are not in any of the obtained subSCCs
(included those initially removed) must be in $T$ as they could be
removed in the termination proof, and are the selected set of
transitions that belong to the corresponding $T_i$ depending on where
is the source location. Fig.~\ref{TSplit1} shows a first split if we
start removing $\tau_1$, as we obtain $C_0$ and $C_1$ and
$T_0=\{\tau_1\}$ and $T_1=\{\tau_2,\tau_7\}$. After this, it is easy
to see that we have conditions~\ref{union}--\ref{dag}. However it may
happen that condition~\ref{sametargettoC0} does not hold, as it is the
case in the example since $\tau_2$ and $\tau_7$ have different target
locations in $C_0$. Then, as shown in Fig.~\ref{TSplit2} we can join
some components until the condition holds again. In this case we join
$C_0$ and $C_1$ into a single component $C_0$ and $T_0$ contains only
$\tau_7$.
\cbend

In what follows, if $C_0, \ldots, C_n$ and
$T_0, \ldots, T_n$ is $\m{split}(C)$ then we define
$\textsf{split-exits}(C_i) = T_i$ and $\textsf{split-entries}(C_i)$ to
all transitions in $T_0,\ldots,T_n$ with target location in $C_i$. We
 call split-entry locations of $C_i$ to the set of target locations
of $\textsf{split-entries}(C_i)$ and split-exit locations of $C_i$ to the
set of source locations of $\textsf{split-exits}(C_i)$.
The following recursive procedure soundly transforms a given component
 with a termination proof only containing linear ranking
functions into a non-cycling
set (forming a tree-like structure) of hierarchically connected loop-nested SCCs (with
the same single entry and exit location) all of them being bounded by
a linear ranking function.


\begin{definition}[transformation to linearly-bounded
  hierarchically loop-nested SCCs]\label{transf}
  Let $C$ be a terminating SCC with $\m{Proof}(C)$ and single entry
  and exit location $e$. Procedure $\m{nestedLoopTrans}(C)$ transforms
  $C$ by first computing $C_0,\ldots,C_n$ and $T_0,\ldots, T_n$ with
  $\m{split}(C)$. If $n=0$ and $C_0$ is a single location, then return
  $C$. Otherwise we perform the following steps: \secbeg
  \secbeg \begin {enumerate} 
\item \label{buildproofs}
    Build all $\m{Proof}(C_i)$ from $\m{Proof}(C)$ following Def.~\ref{def-lexproof} for
    all non-trivial SCC $C_i$. Note that some $C_i$ can include more
    than one component in $\m{Proof}(C)$ and some transitions in $C^R$.
  \item \label{single} Clone all $C_i$ (and $T_i$ and $\m{Proof}(C_i)$) with
    $i>0$ such that $C_i$ has more than one split-entry location. After this, all components have a single split-entry location. \cbstart Then
    apply $moveSourceLocation$ to the resulting components (including the cloned ones) and to $C_0$ to move all transitions in $T_i$ (maybe cloned) with source different from the single split-entry location of $C_i$. \cbend
    After this, we have components $C'_0,\ldots,C'_m$ and
    $T'_0,\ldots, T'_m$, with all $C'_i$ with $i\geq 0$ having a
    single location as both split-entry and split-exit.
  Let $s$ be such location of $C'_0$,
    which may be different from $e$.
  \item \label{looplocations}
    Let $f_i$ be a fresh location if $C'_i$ is a non-trivial SCC or
    the single location in $C'_i$ otherwise.
  \item \label{firstloopout} If $C'_0$ is a trivial SCC only including
    $e$, add all entries of $C$ to the transformation. Otherwise clone
    $C'_0$ obtaining $C''_0$ with a mapping $\mu$ (from old locations
    to fresh locations) and $\m{Proof}(C''_0)$. For every entry
    transition $\langle o,\rho,e\rangle$ in $C$ add an entry
    transition $\langle o,\rho,\mu(e)\rangle$ to $C''_0$ and if $e=s$
    a transition $\langle o,\rho,f_0\rangle$ to the transformation.
    Then, for every transition $\langle s,\rho,t\rangle$ in $T'_0$, add
    a transition $\langle\mu(s),\rho',f_0\rangle$ as exit in $C''_0$
    and for every exit $\langle e,\rho,o\rangle$ of $C$ add as exit in
    $C''_0$ a transition $\langle\mu(e),\rho',f_0\rangle$ if $e=s$ and
    $\langle\mu(e),\rho,o\rangle$ otherwise (where, in all
    cases, $\rho'$ does not include any of the conjuncts with primed
    variables of $\rho$).  If $e\not=s$ then apply $exitToentry$ to
    the resulting $C''_0$ considering that $\mu(e)$ is the entry
    location. Finally, compute $nestedLoopTrans(C''_0)$.
  \item \label{innerloops}
    Replace every transition in $T'_i$ of the form $\langle
    s,\rho,t\rangle$ by $\langle f_i,\rho,t\rangle$ and $\langle
    s,\rho',f_i\rangle$, where $\rho'$ does not include any of the
    conjuncts with primed variables. Note that these transitions are
    new entries and exits of $C'_0,\ldots,C'_m$.
    
  \item \label{exitsouter} Add all exits $\langle e,\rho,o\rangle$ of
    $C$ to $C'_0$ as exit transitions. If $e\not=s$ then apply
    $exitToentry$ to the resulting $C'_0$. Then, replace every
    new exit transition $\langle s,\rho,o\rangle$ of $C'_0$
    by $\langle s,\rho',f_0\rangle$ and add $\langle
    f_0,\rho,o\rangle$ as exit of the transformation of $C$, where,
    again, $\rho'$ is $\rho$ without primed variables.
    
  \item  Compute $\m{nestedLoopTrans}(C'_i)$ on the resulting $C'_i$ for all $i\ge 0$, and add the result to the transformation of $C$.
    
  \end{enumerate}
\end{definition}\secbeg
Intuitively, the steps of the transformation can be understood as
follows. After computing the $split(C)$, step~\ref{buildproofs} builds
the termination proofs associated to the chosen sub-components and
transitions. Then, in step~\ref{single} we turn the components into
components with a single split-entry and split-exit location. \cbstart
For
instance, Fig.~\ref{TSMove} shows the result of applying this step to
the split given in Fig.~\ref{TSplit2}. In this case, we do not need to
clone any component but, as can be seen, we apply $moveSourceLocation$
to $C_0$, $\{\tau_7\}$ and the split-entry location $l_1$ (in
Fig.~\ref{TSplit2}) since the split-entry of $C_0$ is $l_1$ and the
split-exit of $C_0$ is $l_2$. After the step the split-entry and the
split-exit $s$ of $C'_0$ is $l_1$, which is different from the entry
location $e$ which is $l_0$. Note also that $moveSourceLocation$ has changed the entry transition to $C'_0$ adding $n_1'=1$ (which is now the new version of the entry to $C$).
Fig.~\ref{TScomplex}, shows the result of
applying the transformation steps to the $C'_0$ and $T'_0$ in
Fig.~\ref{TSMove} but without applying $\m{nestedLoopTrans}$
recursively. In step~\ref{looplocations}, we define the locations that
are used to express the outer loop (i.e. the loop of all sub-SCCs
$C'_i$) of the transformation. This is $f_0$ in
Fig.~\ref{TScomplex}. In step~\ref{firstloopout} we connect the loop
with the entries of the component. This step is crucial as it includes
an initial use of the first sub-SCC $C'_0$, before entering the outer
loop. The reason for that is that there must be paths in the original
$C$ that run some transitions in $C'_0$ before running any of the
transitions in $T'_0$, which are used as soon as we enter the main
loop. In Fig.~\ref{TScomplex}, we can see the resulting $C''_0$, which
is the result of first cloning $C'_0$, and then, since $l'_0=e\not=s=l'_1$, we add a transition from $l'_1$ to $f_0$ and apply $exitToentry$
to move this transition to $l'_0$. In step~\ref{innerloops}, we connect the
sub-SCCs using the locations $f_i$ to create the outer loop (which is represented by the connections of $C'_0$ to $f_0$ in the right-hand-side of Fig.~\ref{TScomplex}), and
in~\ref{exitsouter} we introduce the needed exit transitions. For simplicity, Fig.~\ref{TScomplex} does not include the exits, but they would be leaving from $l_0=e\not=s=l_1$, and hence $exitToentry$ is applied to move them to $l_1$ and then connected to $f_0$. Finally, 
in the last step we apply the transformation recursively. To further illustrate how the transformation works, the following example shows the complete application of the transformation to our running example. \cbend
\begin{figure}[t]\secbeg
  \centering\secbeg
  \begin{tikzpicture}[auto,>=latex,scale=.7,transform shape]
\tikzstyle{location}=[draw,circle]

\node[location] (0)  {$l_0$};
\node[location] (1)  [right of = 0, node distance = 4.5cm] {$f_0$};
\node[location] (2)  [above of = 1, node distance = 4.4cm] {$l_2$};
\node[location] (3)  [above left = 2cm and 2.5cm of 1] {$\widehat{l_1}$};
\node[location] (11) [below of = 1, node distance = 4cm] {$l_{1}'$};
\node[location] (12) [right of = 1, node distance = 7.5cm] {$l_{1}''$};

\path[->] (0) edge node {$\tau_2': z \geq 1$} (1);

\path[->] (0) edge [out=270,in=186] node [swap] {$\tau_3':z \geq 1$} (11);

\path[->] (0) edge node [near start, xshift=1mm] {$\tau_1': z\leq 0$} (3);

\path[->] (11) edge [loop below] node [xshift=2.7cm, yshift=5mm] {$
    \tau_{16}': x\geq y, z \geq 1, \m{undf1} \leq 0, y' = y + 1
$} (11);

\path[->] (11) edge [bend right, out=-80, in=-150, looseness=1.3] node [swap, yshift=-1.5cm] {$
  \begin{aligned}
    \tau_{15}':\; & x\geq y, z \geq 1, \m{undf1} > 0\\
    & \m{undf2} < x + z
  \end{aligned}
$} (1);

\path[->] (11) edge [bend right=10] node [swap, xshift=-0.5mm, yshift=-5mm] {$
  \begin{aligned}
    &\tau_{14}':\\
    &z\leq 0
  \end{aligned}
$} (1);

\path[->] (11) edge [bend right, out=80, in=150, looseness=1.3] node [near end, xshift=-1mm, yshift=-9mm] {$
  \begin{aligned}
    \tau_{12}':\; & x\geq y, \m{undf1} > 0\\
    & \m{undf2} \geq x + z
  \end{aligned}
$} (1);

\path[->] (11) edge [bend left=10] node [yshift=-5mm] {$
  \begin{aligned}
    \tau_{13}': &\\
    x < y &
  \end{aligned}
$} (1);

\path[->] (1) edge [out=55, in=125] node [midway, xshift=1.5cm, yshift=-0.4mm] {$
  \begin{aligned}
    \tau_7':\; & x\geq y, z \geq 1,\m{undf1} > 0\\
    & \m{undf2} < x + z\\
    & x' = \m{undf2}\\
    & z' = z - 1
  \end{aligned}
$} (12);

\path[->] (12) edge [loop right] node [xshift=-1cm, yshift=-1.4cm] {$
  \begin{aligned}
    \tau_{21}':\; & x\geq y\\
    & z \geq 1\\
    & \m{undf1} \leq 0\\
    & y' = y + 1
  \end{aligned}
$} (12);

\path[->] (12) edge [bend left=40] node {$
  \begin{aligned}
    \tau_{20}':\; & x\geq y, z \geq 1, \m{undf1} > 0\\
    & \m{undf2} < x + z\\
  \end{aligned}
$} (1);

\path[->] (12) edge node {$\tau_{18}': z\leq 0$} (1);

\path[->] (12) edge [bend right=35] node [yshift=-1mm] {$
  \begin{aligned}
    \tau_{17}':\; & x\geq y, \m{undf1} > 0\\
    & \m{undf2} \geq x + z\\
  \end{aligned}
$} (1);

\path[->] (12) edge [bend left=20] node {$\tau_{19}': x < y$} (1);

\path[->] (1) edge node [near end, xshift=3mm, yshift=-1.5mm] {$ \tau_4': z\leq 0$} (3);

\path[->] (1) edge [bend right=10] node [swap, xshift=-1.5mm, yshift=8mm]{$
  \begin{aligned}
    \tau_6':\; & x\geq y,\m{undf1} > 0\\
    & \m{undf2} \geq x + z\\
    & x' = \m{undf2}\\
    & z' = z - 1
  \end{aligned}
$} (2);

\path[->] (1) edge [bend left=10] node [yshift=-2mm] {$\tau_5': x < y$} (2);

\path[->] (3) edge [loop left] node [yshift=8mm] {$
  \begin{aligned}
    \tau_8':\; & x\geq y\\
    & z \leq 0\\
    & \m{undf1} > 0\\
    & \m{undf2} < x + z\\
    & x' = \m{undf2}\\
    & z' = z - 1
  \end{aligned}
$} (3);

\path[->] (3) edge [loop above] node [xshift=-3mm, yshift=1mm] {$
  \begin{aligned}
    \tau_9':\; & x\geq y\\
    & z \leq 0\\
    & \m{undf1} \leq 0\\
    & y' = y + 1
  \end{aligned}
$} (3);

\path[->] (3) edge [bend left=11] node [very near end, xshift=10mm] {$
  \begin{aligned}
    \tau_{10}':\; & x\geq y, \m{undf1} > 0\\
    & \m{undf2} \geq x + z\\
    & x' = \m{undf2}\\
    & z' = z - 1
  \end{aligned}
$} (2);

\path[->] (3) edge [bend right=11] node [swap, near start, xshift=-3mm] {$\tau_{11}': x < y$} (2);

\end{tikzpicture}
%
%
%
%
%
%
%
%
%
%
\secbeg\secbeg
\caption{
Transformed CRS} \label{TSrunning} \vspace{-0.4cm}
\end{figure}

\secbeg\secbeg\begin{example}
  \label{runningtransLex}
  Let us show how it works starting from the transformed graph in
  Fig.~\ref{fig:unfolding1} and with the termination proofs for each
  phase given in Ex.~\ref{runningproof}. The resulting transition
  system is shown in Fig.~\ref{TSrunning}. Its key feature is that it
  is ready to generate a linearly-bounded CRS in next section.  There
  are two SCCs in Fig.~\ref{fig:unfolding1}. The SCC that cycles in
  the location $\widehat{l_1}$ is proved with a single ranking
  function where all transitions are removed, and hence our
  transformation does not change anything, since $C_0$ is the location
  $\widehat{l_1}$ and $T'_0=T_0$ contains both transitions. The SCC
  that cycles in location $l_1$, needs a lexicographic combination of
  two ranking functions, with each component removing one transition,
  firstly removing $\tau_1\land z\geq 1$ and secondly
  $\tau_2\land z\geq 1$. Then $\m{split}$ gives
  $T_0=\{\tau_1\land z\geq 1\}$ and $C_0$ is the SCC including location $l_1$ and
  transition $\tau_2\land z\geq 1$.  Therefore, first of all we compute
  $\m{Proof}(C_0)$ according to step~\ref{buildproofs} of
  Def.~\ref{transf}. Step~\ref{single} does not change anything, since
  we have a single component $C'_0=C_0$ with a single
  location. Step~\ref{looplocations} delivers a fresh location
  $f_0$. Next, since $C'_0 = C_0$ is a non-trivial SCC, we clone it to
  $C''_0$ in step~\ref{firstloopout}. This new SCC corresponds to
  location $l_1'$ and transition $\tau_{16}' = \tau_2\land z\geq 1$ in
  Fig.~\ref{TSrunning}. Transition $\tau_3'$ is the entry added to
  $C_0''$. A transition $\tau_2'=z\geq 1$ entering $f_0$ is
  also added since in this case $e=s$.  Transitions $\tau_{12}'$,
  $\tau_{13}'$, $\tau_{14}'$ and $\tau_{15}'$ are the transitions
  added as exits. Note that $\tau_3'$ expresses the same transition
  relation as the entry $\tau_0 \land z'\geq 1$ of $l_1$, since
  $\tau_0 = \m{true}$. Transition $\tau_{15}'$ is the same as the
  transition $\tau_1\land z\geq 1$ in $T_0$ without conjuncts with
  primed variables, while $\tau_{12}'$, $\tau_{13}'$ and $\tau_{14}'$
  express the same transition relation as the original exits of $l_1$
  with $\tau_3$, $\tau_4$ and $z\leq 0$, respectively, except for the
  conjuncts with primed variables again. In step~\ref{innerloops}, the
  only transition in $T_0$ is unfolded into the transitions $\tau_7'$
  and $\tau_{20}'$ of Fig.~\ref{TSrunning}. Step~\ref{exitsouter} adds
  transitions $\tau_{17}'$, $\tau_{18}'$ and $\tau_{19}'$,
  corresponding to the exit transitions of $l_1$ but without conjuncts
  with primed variables, as well as transitions $\tau_4'$, $\tau_5'$
  and $\tau_6'$ as exits of the transformation of $C$. Finally, notice
  that $l_1$ has been renamed to $l_1''$ to avoid confusion with the
  original location, and $\tau_{21}'$ corresponds to
  $\tau_2\land z\geq 1$.  Note that all recursive calls to
  \emph{nestedLoopTrans} trivially terminate in this example.
\end{example}
The transformation provided in this section is sound for resource analysis.
\begin{theorem}[soundness and linearly-bounded]\label{th1}
  Given a component $C$ with $\m{Proof}(C)$, then every SCC in
  $nestedLoopTrans(C)$ is a \emph{linearly-bounded hierarchically loop-nested
    transition system}, and for every path $\pi$ from an initial
  location $s$ to a final location $t$ in $C$, there is a path $\pi'$ from $s$ to $t$ in
  $nestedLoopTrans(C)$ with $\#(\pi') \geq \#(\pi)$, where $\#(\pi)$ is the number of operations involved in $\pi$.

\end{theorem}

\secbeg
\secbeg

\secbeg
\secbeg
\section{Linearly-Bounded Cost Relation Systems}

A CRS is a set of cost equations of the form \cbstart
$c(\overline{x})=1+c_1(\overline{x_1})+\ldots+c_n(\overline{x_n})
\{\constraints\}$,
where the \emph{constraints} $\constraints$ \cbend define the applicability conditions
for the equation and state size relations among $\overline{x}$,
$\overline{x_1},\ldots,\overline{x_n}$. As
stated in Sec.~\ref{sec:introduction}, w.l.o.g., we always accumulate
a constant unitary cost. The set of cost equations for $c(\overline{x})$
defines the (possibly non-deterministic) cost function $c$.  Even if
the input language from which the CRS are produced is deterministic,
due to the loss of information implicit to static analysis (e.g.,
when \emph{undef} variables appear), the
associated CRS will typically be non-deterministic. 
\cbstart 
CRS can be considered  as
constraint logic programs over integers that accumulate costs, e.g., the above
equation can be written as
the clause $c(\overline{X},Co)\textnormal{~:-~}\constraints,~c_1(\overline{X_1},Co_1),~\ldots,~$ $c_n(\overline{X_n},Co_n),~Co\textnormal{~\#=~}1+Co_1+\ldots+Co_n$
(see also the CRS in Fig.~\ref{running}).  
\cbend 
%

The following definition presents the generation of a CRS from a TS
with possible multiple nested loops. 
As explained in
Sec.~\ref{sec:introduction}, we  assume   that 
a language-specific size analysis has been already applied such that
entry locations of SCC/sub-SCC in the TS are annotated with
$\sizean(l, \langle\overline{x}\rangle, \langle\overline{x'}\rangle)$:
\emph{the size relations between the values of the variables when
reaching ($\overline{x}$) and leaving ($\overline{x'}$) a location
$l$}. For example, for the TS in Fig.~\ref{TSrunning}, the size
analysis of location $l''_1$ will infer the relations
$\sizean(l''_{1}, \langle x,y,z\rangle, \langle x',y',z'\rangle) =
\{x=x', y' \geq y, z = z'\}$
as any path $l''_{1} \to^* l''_{1}$ using $\tau'_{21}$ will not modify $x$ and $z$, and
$y$ can only increase. Size analyses for various languages can be
found e.g. at
\cite{AlbertAGPZ07-short,DBLP:journals/tplp/SerranoLBH13,DBLP:journals/toplas/BrockschmidtE0F16}.


\begin{figure}[t]
\noindent
\begin{minipage}[t]{.5\textwidth}
\[
\begin{array}{l}
\hspace*{-.4cm}\circled{1}~c_{l_0}(x,y,z) = 1 + c_{f_0}(x,y,z) ~\{z \geq 1\}\\
\hspace*{-.4cm}\circled{2}~c_{l_0}(x,y,z) = 1 + c_{l'_1}(x,y,z) ~\{z \geq 1\}\\
\hspace*{-.4cm}\circled{3}~c_{l_0}(x,y,z) = 1 + c_{\widehat{l_1}}(x,y,z) ~\{z \leq 0\}\\ 

\hspace*{-.4cm}\circled{4}~c_{l_2}(x,y,z) = 1 ~\{\} \\ 

\hspace*{-.4cm}\circled{5}~c_{f_0}(x,y,z) = 1 + c_{l''_1}(x_1,y_1,z_1) + c_{f_0}(x_2,y_2,z_2) \\
~~\{
x_1 = x_2, y_2 \geq y_1, z_1 = z_2, x\geq y, z\geq 1, u_1 > 0, \\
~~~~u_2 < x+z,x_1 = u_2, y_1=y,z_1 = z-1\}\\
\hspace*{-.4cm}\circled{6}~c_{f_0}(x,y,z) = 1 + c_{\widehat{l_1}}(x,y,z) ~\{z \leq 0\} \\
\hspace*{-.4cm}\circled{7}~c_{f_0}(x,y,z) = 1 + c_{l_2}(x,y,z) ~\{x < y\} \\
\end{array}
\]
\end{minipage}
\begin{minipage}[t]{.5\textwidth}
{\small
\[
\begin{array}{l}
\hspace*{-.4cm}\circled{8}~c_{f_0}(x,y,z) = 1 + c_{l_2}(x',y',z') ~\{x \geq y, u_1 > 0, \\
~~u_2 \geq x+z,
x' = u_2, y'=y, z' = z-1\} \\ 

\hspace*{-.4cm}\circled{9}~c_{l''_1}(x,y,z) = 1 + c_{l''_1}(x',y',z') ~\{x \geq y, z \geq 1, \\
~~u_1 \leq 0, x'=x, y'=y+1, z'=z\}\\
\hspace*{-.4cm}\circled{10}~c_{l''_1}(x,y,z) = 1 ~\{x \geq y, u_1 > 0, u_2 \geq x+z\}\\
\hspace*{-.4cm}\circled{11}~c_{l''_1}(x,y,z) = 1 ~\{z \leq 0\}\\
\hspace*{-.4cm}\circled{12}~c_{l''_1}(x,y,z) = 1 ~\{x < y\}\\
\hspace*{-.4cm}\circled{13}~c_{l''_1}(x,y,z) = 1 ~\{x \geq y, z \geq 1, u_1 > 0, u_2 < x+z\}\\
....
\end{array}
\]}
\end{minipage}
\caption{Fragment of Linearly-bounded CRS obtained from transformed
  TS }\label{fig:finalCRS}
\vspace{-0.4cm}
\end{figure}




\begin{definition}[linearly-bounded CRS]\label{crs}
  \cbstart Given a linearly-bounded hierarchically-loop-nested TS $\its$, let $G$ be the set of locations in $\its$,
  $\overline{x}$ be the variables involved in $\its$ and $\sccproc(\locs)$
  the list of the SCCs in $\its$
  considering only the locations in the set $\locs$. Let $\entry(\locs)$
  denote the entry location of a set of locations $\locs$ (i.e., the only
  location receiving transitions from outside $\locs$).  The
  LB-CRS for \its{} is made up of the cost equations
  generated by $\transITS(G)$ that, for every SCC $\locs \in \sccproc(G)$,
  proceeds as follows: \cbend

\begin{itemize}
\item If $\cbstart \locs \cbend = \{l_o\}$, i.e., the SCC contains only one location $l_o$, then every transition $l_o \totau{\tau} l_d$ generates one cost equation $c_{l_o}(\overline{x}) = 1 + c_{l_d}(\overline{x}')~\{\tau\}$. If there are no transitions from $l_o$, a dummy equation $c_{l_o}(\overline{x}) = 1\{\}$ is generated for uniformity.
\item If $\cbstart |\locs| \cbend > 1$ and $\entry(\locs) = l_o$, as the entry location is
  part of the principal loop in $\cbstart \locs \cbend$, we remove it to detect and, transitively,
  translate  the remaining components. Every
   $D_i \in \sccproc(\cbstart \locs \cbend \setminus \{l_o\})$
  is translated by
  \cbstart $\transInner(D_i)$---defined below---and every cycle in the
  \emph{component graph} (each SCC $D_i$ is condensed into a single
  vertex $d_i$) starting from $l_o$, i.e., cycles paths of the form
  $l_o \totau{\tau} d_1 \to d_2 \ldots \to d_j \to l_o$, generates an equation \cbend
(where $n_i = \entry(D_i)$):
\[\begin{array}{lll} 
c_{l_o}(\overline{x}) & = & 1 + c_{n_1}(\overline{x'}) + c_{n_2}(\overline{x^{(2)}}) + \ldots + c_{n_j}(\overline{x^{(j)}}) + c_{l_o}(\overline{x^{(j+1)}}) \\
& &  \{\tau\} \cup \sizean(l_{n_1}, \lr{\overline{x'}},\lr{\overline{x^{(2)}}}) \cup \ldots \cup \sizean(l_{n_j}, \lr{\overline{x^{(j)}}},\lr{\overline{x^{(j+1)}}})
\end{array}
\]
Every outgoing transition $l_o \totau{\tau} l_k$ ($l_k \notin \cbstart\locs\cbend$) generates an equation $c_{l_o}(\overline{x}) = 1 + c_{l_k}(\overline{x'}) \{\tau\}$.
\end{itemize}
$\transInner{}$ proceeds as $\transITS{}$ with one difference: in both cases \cbstart ($\locs=\{l_o\}$ and $|\locs|>1$) \cbend the outgoing transitions $l_o \totau{\tau} l_k$ generate cost equations $c_{l_0}(\overline{x}) = 1 \{\tau\}$, i.e., without any call.
\end{definition}
Let us give the intuition behind the above definition. For each location
$l$ in the TS, we produce a corresponding cost function $c_l$ that
captures its cost, and every transition produces an equation.  The
labels of the transitions become the constraints of the CRS.  The
equation for the cycle in the component graph collects the costs of all
the sequential components and finishes with a recursive call to
express the loop. The size analysis allows us to track the changes in
the variables after every function call in order to express the cost
in terms of the initial parameter values. \cbstart Note that the constraints
of the transitions $d_i \to d_{i+1}$ and $d_j \to l_o$ are not needed in the equation because they
have been already used when generating the equations for every SCC
$D_i$ recursively. \cbend Finally, as $\transInner{}$ is applied to
components in an inner loop, the flow represented by these outgoing
transitions is incorporated in the cost equation of the outer loop and
no function call is needed.

\begin{example}\label{ex:gencrs}
\cbstart
In
  the transformed TS from Fig.~\ref{TSrunning} there are 5 SCCs:
  $\{l_0\}$, $\{l_2\}$, $\{l'_1\}$, $\{\widehat{l_1}\}$ and
  $\{f_0, l''_1\}$. The first four are unitary, thus they generate
  equations directly. \cbstart For example in $l_0$, the transitions $l_0 \to f_0$, $l_0 \to l'_1$, and $l_0 \to \widehat{l_1}$ generate equations with calls to the corresponding cost function (see equations \#1--3 in Fig.~\ref{fig:finalCRS}). On the other hand, $l_2$ has no outgoing transition hence it creates a dummy equation for $c_{l_2}$ (\#4). Considering the non-unitary $\{f_0, l''_1\}$, the
  only SCC after removing the entry location $f_0$ is $\{l''_{1}\}$, thus $\transInner(\{l''_1\})$ generates 5 equations for $c_{l''_1}$: $\tau'_{21}$ creates the recursive equation (\#9) and $\tau'_{17}$, $\tau'_{18}$, $\tau'_{19}$, and $\tau'_{20}$ generate 4 equations without any call (\#10--13).
  As $f_0$ is the entry location, its outgoing transitions generate
  the non-recursive equations of $c_{f_0}$ that invoke $c_{\widehat{l_1}}$ and $c_{l_2}$ (\#6--8). In the component graph $l''_1$ is condensed into $d_{l''_1}$ by removing the transition $\tau'_{22}$, hence there is only one cycle 
  $f_0 \totau{} d_{l''_1} \to f_0$ that generates the
  recursive equation of $c_{f_0}$ (\#5).
  The size analysis relates the input
  and output values after invoking function $c_{l''_1}$ in this equation
  ($x_1 = x_2, y_2 \geq y_1, z_1 = z_2$), which allows tracking the
  changes from the initial values $x,y,z$ to the ones used in the
  recursive call $x_2,y_2,z_2$ when solving the CRS. \cbend This CRS is
  solvable because, thanks to the transformation of the TS, all
  cost functions have now a linear ranking function that bounds the
  number of calls. The solver can hence use that information to generate
  the overall cost. Concretely, $c_{\widehat{l_1}}$, $c_{l'_1}$, and $c_{l''_1}$ are
  invoked ${x-y}$ times; $c_{f_0}$ is invoked ${z}$ times; and
  $c_{l_0}$ and $c_{l_2}$ are invoked only once. This contrasts with the
  original CRS in Fig.~\ref{running}, where the
   lexicographic ranking function $\langle {z},{x-y}\rangle$
  cannot be used by the backend solvers to compute a
  loop bound.
\cbend  
\end{example}
\secbeg
The next corollary easily follows  from the soundness of the TS
transformation in Th.~\ref{th1}. 

\begin{corollary}[soundness of linearly-bounded CRS]
\label{soundCRS}
 Let  $N=nestedLoopTrans(C)$ be the \emph{hierarchically loop-nested
    transition system} obtained from a component $C$ with
  $\m{Proof}(C)$. The CRS obtained from $N$ applying
  Def.~\ref{crs} 
  soundly overapproximates the cost
  of $C$ (for the considered cost model), and all its functions are linearly bounded.
\end{corollary}

\secbeg
\secbeg
\secbeg
\secbeg
\secbeg
\secbeg
\subsection{Embedding the Ranking Functions from Termination Proofs within CRS}\label{sec:incl-rank-funct}
\secbeg

CRS solving ---step (2) of resource analysis---
requires finding ranking functions for all recursive cost functions
(i.e. cycles)  to bound the number of iterations they might perform. As the
termination analyzer must have already found ranking functions for all
cycles, it is desirable to pass
this information to the CRS solver (e.g., the resource analyzer might
implement less powerful algorithms to find ranking
functions). However, existing solvers are not prepared to receive this
information.
Our proposal does not require implementing any
 extension to the existing solvers. We can embed the
constraints that define the ranking functions within the CRS as
follows.

\begin{definition}[CRS with ranking functions]
We assume that every location $l$ in the TS is annotated with the linear ranking function contained in the termination proof.
The main idea is to add a new parameter to the cost function $c_l$ representing the ranking function, which is bound in the initial call and decreases in every recursive call. Therefore, the generation of the cost equations
is as in Def.~\ref{crs} with the following differences:
\secbeg\secbeg\secbeg\secbeg\begin{itemize}
\item Cost functions $c_l$ (except those for the initial location $l_0$) are extended with one additional parameter $r$ representing the ranking function: $c_l(\overline{x},r)$.
\item Cost equations invoking a cost function with ranking function $\mi{rf}$ bound the extra parameter: $\{r = \mi{rf}\}$.
\item Cost equations with recursive calls are extended to express that the extra parameter containing the ranking function is positive and strictly decreasing: $\{r\geq 0, r' < r\}$.
\end{itemize}
\end{definition}\secbeg \secbeg
\begin{example} 
  Let us explain the above definition using our running example.  In
  the transformed TS of Fig.~\ref{TSrunning}, the termination
  analyzer detects that $f_0$, $l''_1$, and $l_{\widehat{l_1}}$ have ranking functions $z$, $x-y$, and $x-y$ resp.
  Then, the cost equations of $l_0$, $f_0$, and $l''_{1}$ will be modified as follows (we show only a fragment):
\secbeg  
\[ \small
\begin{array}{ll@{~}c@{~}l}
\circled{1'}&c_{l_0}(x,y,z) & = & 1 + c_{f_0}(x,y,z,r) ~\{z \geq 1, \mathbf{r=z}	\}\\[-.15cm]
& & \ldots\\[-.12cm]
\circled{5'}& c_{f_0}(x,y,z,r) & = & 1 + c_{l''_1}(x_1,y_1,z_1,r_1) + c_{f_0}(x_2,y_2,z_2,r_2) ~\{
x_1 = x_2, y_2 \geq y_1, z_1 = z_2, x\geq y,z\geq 1,\\
& & &  u_1 > 0, u_2 < x+z,x_1 = u_2, y_1=y,z_1 = z-1, \mathbf{r\geq 0, r_1 = x_1 - y_1, r_2 < r}\}\\
\circled{6'} & c_{f_0}(x,y,z,r)      & = & 1 + c_{\widehat{l_1}}(x,y,z,r') ~\{z \leq 1, \mathbf{r'=x-y}\}\\[-.15cm]
& & \ldots\\[-.12cm]
\circled{9'} & c_{l''_1}(x,y,z, r)   & = & 1 + c_{l''_1}(x_1,y_1,z_1,r_1) ~\{x \geq y, z \geq 1, u_1 \leq 0, x'=x, y'=y+1, z'=z, \mathbf{r\geq 0, r_1 < r}\}\\
\circled{13'} & c_{l''_1}(x,y,z,r)   & = & 1 ~\{x \geq y, z \geq 1, u_1 > 0, u_2 < x+z\}\\[-.15cm]
& & \ldots\\[-.12cm]
\end{array}
\]
\cbstart
The remarked constraints in the equations represent the changes. In equations $1'$ and $6'$ the ranking functions of $f_0$ ($z$) and $l_{\widehat{l_1}}$ ($x-y$) are bound. In the recursive equations $5'$ and $9'$, the extra parameter is set to positive and decreasing. Additionally, in $5'$ the extra parameter $r_1$ of $l''_1$ is bound to $x_1-y_1$. Finally, cost equations without invocations (as equation $13'$) are not modified. Note that $c_{l_0}$ (equation $1'$) is not extended with any parameter because it is the entry location of the program.\cbend
\end{example}

\secbeg
\secbeg
\secbeg
\secbeg
\secbeg
\secbeg
\subsection{Extension to Conditional Upper Bounds}\label{sec:extens-gener-cond}
\secbeg

The termination analysis we use
\cite{DBLP:conf/tacas/BorrallerasBLOR17} is able to infer
preconditions $\m{Pre}$ under which the program terminates when it cannot
 prove termination for all inputs. Such preconditions
$\m{Pre}$ may also be valid for the upper bounds. As in
Sec.~\ref{sec:incl-rank-funct}, the idea is to embed
  $Pre$ into the CRS (and enable a flag
 \textsf{cond=on}) so that an unconditional solver can be used and, as
 preconditions are assumed, a \emph{conditional upper bound} $U$ can now be
found. Then, when reporting the results, if \textsf{cond=on}, we
output that
$U$ is an upper bound if preconditions $\m{Pre}$ hold.

\begin{definition}[conditional CRS]
Let $G$ be the set of locations in the TS and $l_0$ be its
entry location. 
Then the cost equations of the conditional CRS are $\transITS(G)$ plus the additional equation $c_{l_0}^e(\overline{x}) = 1 + c_{l_0}(\overline{x}) \{\m{Pre}\}$. In this case, the upper bound obtained for $c_{l_0}^e$ is the valid upper bound for $c_{l_0}$ under the conditions in $\m{Pre}$.

\end{definition}


\secbeg
\secbeg
\secbeg
\section{Implementation and Experimental Evaluation}

\begin{table}
\cbstart
{\footnotesize
 \begin{tabular}{rr@{}r@{\hskip -3pt}c@{\hskip -3pt}r@{}r@{\hskip -3pt}c@{\hskip -3pt}r@{}r@{}r@{}r}
 \toprule
 & \textbf{\pubsc}
 & \textbf{\maxcorepubss}
 & $\vline$
 & \textbf{\coflococ}
 & \textbf{\maxcorecoflocos}
 & $\vline$  
 & \textbf{\aprove}
 & \textbf{\coflococ}
 & \textbf{\loopus}
 & \textbf{\maxcorecoflocos} \\
 \midrule
 
 \multirow{2}{*}{\code{Solved}} & $158$ & $280$ & $\vline$ & 
 $288$ & $311$ & $\vline$ & $278$ & $288$ & $239$ & $311$ \\
 & $(32.6\%)$ & $(57.9\%)$ & $\vline$ & 
 $(59.5\%)$ & $(64.3\%)$ & $\vline$ & $(57.4\%)$ & $(59.5\%)$ & $(49.4\%)$ & $(64.3\%)$ \\
 
 \code{Only} & $6$ & $128$ &  $\vline$ & 
 $21$ & $44$ & $\vline$ & $2$ & $5$ & $3$ & $33$\\
 
 \code{Best} & $20$ & $128$ &  $\vline$ & 
 $32$ & $47$ & $\vline$ & $2$ & $7$ & $9$ & $38$\\
 
 \code{Score} & $316$ & $546$ & $\vline$ & 
 $573$ & $611$ &$\vline$ & $1075$ & $1147$ & $946$ & $1228$\\
 
 \multirow{2}{*}{\code{Time(s)}} & \multirow{ 2}{*}{$836.8$} & {\scriptsize gen:~}$837.5$ & $\vline$ & 
 \multirow{ 2}{*}{$1175.6$} & {\scriptsize gen:~}$838.6$ & $\vline$ & \multirow{ 2}{*}{$2350.9$} & \multirow{ 2}{*}{$1175.6$} & \multirow{ 2}{*}{$9.38$} & {\scriptsize gen:~}$838.6$\\
  & & {\scriptsize sol:~}$1574.4$ & $\vline$ & 
  & {\scriptsize sol:~}$1272.3$ & $\vline$ & & & & {\scriptsize sol:~}$1272.3$ 
  \\
 \bottomrule
 \end{tabular}}\cbend
\caption{Experimental results on C programs from TermComp'19 complexity
  competition} \label{tabla}\secbeg\secbeg
\end{table}

Our implementation, \textsf{MaxCore(X)} where \textsf{X} instantiates
the CRS solver, achieves the cooperation of three advanced tools for
complexity and termination analysis: \verymax (winner of
TermComp'19 for C programs) produces the termination proofs, our
implementation generates from them LB-CRS that are fed: (\textsf{X=C})
to \cofloco~\cite{DBLP:phd/dnb/Montoya17}
or (\textsf{X=P}) to \pubs~\cite{AlbertAGP08-short} to produce the
upper bounds. \toolname can be used online from a web interface
\url{https://costa.fdi.ucm.es/maxcore}, where the benchmarks used for
our experiments can also be found. This section evaluates the
effectiveness and efficiency of \toolname by analyzing all C programs from
the TermComp'19 complexity competition, that in total are 484
benchmarks containing also non-terminating programs. Experiments have been performed on an Intel Core i7-4790
at 3.6GHz x 8 and 16GB of memory, running Ubuntu 18.04.
%
 The row \code{Solved} in Table~\ref{tabla}  shows the
number of benchmarks that each system is able to bound. 
The row \code{Only} shows the number of benchmarks that only the
corresponding system can solve and no other system can. The row \code{Best}  counts the times that a system has obtained the best upper bound, and the remaining systems have larger bounds. 
The \code{Score} represents the points obtained by
the systems following the competition rules
\url{http://cbr.uibk.ac.at/competition/rules.php}.
Finally, we show the overall time in seconds in the row
\code{Time(s)}. As in TermComp'19, systems  only have $300$ seconds to analyze every program.
For \toolname's instantiations we show 2 values: the time needed to generate the CRS (\emph{gen}) and the time required to obtain a closed upper bound (\emph{sol}).
Detailed results for every system and benchmark can be found at \url{https://costa.fdi.ucm.es/maxcore/benchmarks/}. 

The left part of Table~\ref{tabla} compares
\pubsc  and \maxcorepubss. {\label{foot}\pubs and \cofloco are CRS solvers. To
  avoid confusion, we use \pubsc and \coflococ for the systems that
  translate C programs to CRS using \code{clang}
  (\url{http://clang.llvm.org/}) and \code{llvm2KITTeL}
  (\url{https://github.com/s-falke/llvm2kittel}), and then use the
  respective CRS solver to obtain an upper bound.}
Unlike \cofloco, \pubs only works with linear size relations, but it
is able to obtain logarithmic upper bounds. As TermComp'19 only
supports polynomial bounds, in the comparisons we have considered
$O(n^k\times \mi{log}^p(n))$ equal to $O(n^{k+1})$ in \code{Best} and
\code{Score}. Since \pubs is a solver that does not perform any
additional analysis on the CRS (unlike \cofloco, which tries to detect
\emph{chains}), this comparison plainly shows the large improvement
that can be only attributed to the proposed generation of our
LB-CRS. Concretely, \maxcorepubss almost doubles the number of
programs solved ($280$ vs. $158$), and there are 128 programs that
\maxcorepubss solves that \pubsc cannot, while only $6$ programs are
uniquely solved by \pubsc. Regarding time, \maxcorepubss is about
three times slower than \pubsc but the gains clearly justify
the additional analysis time.

The central part of Table~\ref{tabla} shows the comparison between
\coflococ and \maxcorecoflocos. Here the
difference is not as large as with \pubs, but it is still important:
\maxcorecoflocos solves $23$ programs more than \coflococ ($311$
vs. $288$), $44$ of them that \coflococ cannot handle. However,
\coflococ solves $21$ programs that \maxcorecoflocos cannot and
obtains better upper bounds in $32$ programs. Since \maxcorecoflocos
uses \verymax to build the termination proof that guides the
generation of the LB-CRS, the system returns $\infty$ if \verymax
cannot find that proof.
This happens in $6$ of these $21$
unsolved programs. 
Moreover, other $5$ programs are not solved because their termination
proof presents some features not yet integrated in the system, but
are planned to be integrated soon. For the remaining 10 unsolved programs,
the termination proofs found by \verymax are too involved, thus leading
to (unnecessarily) more complex CRS that \cofloco cannot handle. Note
that \verymax can find different proofs for a program, and currently
we simply use the first one. In the future, we plan to investigate
on finding the best suited proofs for the solving step.
\maxcorecoflocos has a running time $1.8$ larger than \coflococ, so
the difference is smaller than with \pubs, and clearly it pays
off as well. Comparing \maxcorepubss and \maxcorecoflocos, the latter is about
$300$ seconds faster and obtains better results in all
metrics. Indeed, all programs that can be bound with \maxcorepubss can
be bound with equal or smaller bounds by \maxcorecoflocos except for
one program. Therefore, we have selected \maxcorecoflocos to compare
to the two systems participating in TermComp'19: \cbstart \coflococ, winner of
TermComp'19 for complexity of ITS and C programs, and 
\aprove~\cite{DBLP:conf/rta/GieslTSF04}, a system that implements an
alternative approach which alternates between finding runtime bounds
and finding size bounds
\cite{DBLP:journals/toplas/BrockschmidtE0F16}. Additionally, we have also considered the \loopus 
system~\cite{SZV14:CAV} described in Sec.~\ref{sec:introduction}. 
\cbend
\cbstart The   results of this comparison
appear in the right part of Table~\ref{tabla}, where it can be seen
that \maxcorecoflocos outperforms the other systems in all metrics:
besides number of problems solved, more importantly \maxcorecoflocos
solves $33$ programs that no other system can bound, generates better
bounds in $38$ programs, and obtains $81$ more points than \coflococ, 
$153$ more than \aprove, and $282$ more than \loopus.  Moreover, it is slightly faster than
\aprove, requiring $0.9$ times its running time. 
Finally, note that \loopus is extremely fast compared to the rest of
systems (it requires $0.008$ times the running time of \coflococ, the
second fastest system). The reason is that \loopus relies on
difference logic, a more limited domain for obtaining bounds than the
linear integer arithmetic used in the rest of systems, for 
which
very efficient algorithms exist. \cbend 


\secbeg
\secbeg
\secbeg
\secbeg

\section{Conclusions}

This paper brings the important advances achieved in the field of
termination analysis, where programs featuring complex control flow
can be automatically proven to terminate, to the field of resource
analysis, where there is more limited support for such kind of
complex-flow programs. The success of our approach is the use
of termination proofs as semantic guidance to generate linearly-bounded
CRS that can be fed to an off-the-shelf CRS solver. 
%
Our experimental results on the TermComp'19 benchmarks show that
our tool, \toolname, outperforms the standalone resource analyzers \cofloco,
 \aprove, and \loopus 
significantly both in accuracy, number of problems solved,
and uniquely solved.
As future work, we plan to apply precondition inference techniques~\cite{DBLP:journals/tplp/KafleGGS18} to improve the precision of the termination proof when assertions are provided.

\secbeg\secbeg\secbeg\secbeg\secbeg
\bibliographystyle{acmtrans}
\bibliography{biblio}

\begin{thebibliography}{}

\bibitem[\protect\citeauthoryear{Albert, Arenas, Genaim, and Puebla}{Albert
  et~al\mbox{.}}{2008}]{AlbertAGP08-short}
{\sc Albert, E.}, {\sc Arenas, P.}, {\sc Genaim, S.}, {\sc and} {\sc Puebla,
  G.} 2008.
\newblock Automatic inference of upper bounds for recurrence relations in cost
  analysis.
\newblock In {\em Proc. of SAS 2008}. LNCS, vol. 5079. Springer, 221--237.

\bibitem[\protect\citeauthoryear{Albert, Arenas, Genaim, Puebla, and
  Zanardini}{Albert et~al\mbox{.}}{2007}]{AlbertAGPZ07-short}
{\sc Albert, E.}, {\sc Arenas, P.}, {\sc Genaim, S.}, {\sc Puebla, G.}, {\sc
  and} {\sc Zanardini, D.} 2007.
\newblock {C}ost {A}nalysis of {J}ava {B}ytecode.
\newblock In {\em Proc. of ESOP'07}. LNCS, vol. 4421. Springer, 157--172.

\bibitem[\protect\citeauthoryear{Albert, Correas, Ka~I~Pun, and
  Rom\'an-D\'iez}{Albert et~al\mbox{.}}{2018}]{AlbertCPJR18}
{\sc Albert, E.}, {\sc Correas, J.}, {\sc Ka~I~Pun, E. B.~J.}, {\sc and} {\sc
  Rom\'an-D\'iez, G.} 2018.
\newblock {P}arallel {C}ost {A}nalysis.
\newblock {\em ACM Trans. Comput. Log.\/}~{\em 19,\/}~4, 1--37.

\bibitem[\protect\citeauthoryear{Borralleras, Brockschmidt, Larraz, Oliveras,
  Rodr{\'{\i}}guez{-}Carbonell, and Rubio}{Borralleras
  et~al\mbox{.}}{2017}]{DBLP:conf/tacas/BorrallerasBLOR17}
{\sc Borralleras, C.}, {\sc Brockschmidt, M.}, {\sc Larraz, D.}, {\sc Oliveras,
  A.}, {\sc Rodr{\'{\i}}guez{-}Carbonell, E.}, {\sc and} {\sc Rubio, A.} 2017.
\newblock Proving termination through conditional termination.
\newblock In {\em Proc. {TACAS} 2017}. LNCS, vol. 10205. Springer, 99--117.

\bibitem[\protect\citeauthoryear{Brockschmidt, Emmes, Falke, Fuhs, and
  Giesl}{Brockschmidt
  et~al\mbox{.}}{2016}]{DBLP:journals/toplas/BrockschmidtE0F16}
{\sc Brockschmidt, M.}, {\sc Emmes, F.}, {\sc Falke, S.}, {\sc Fuhs, C.}, {\sc
  and} {\sc Giesl, J.} 2016.
\newblock Analyzing runtime and size complexity of integer programs.
\newblock {\em {ACM} Trans. Program. Lang. Syst.\/}~{\em 38,\/}~4, 13:1--13:50.

\bibitem[\protect\citeauthoryear{Cousot and Halbwachs}{Cousot and
  Halbwachs}{1978}]{DBLP:conf/popl/CousotH78}
{\sc Cousot, P.} {\sc and} {\sc Halbwachs, N.} 1978.
\newblock Automatic discovery of linear restraints among variables of a
  program.
\newblock In {\em Proc. {POPL} 1978}. ACM, 84--96.

\bibitem[\protect\citeauthoryear{Debray and Lin}{Debray and
  Lin}{1993}]{DBLP:journals/toplas/DebrayL93}
{\sc Debray, S.~K.} {\sc and} {\sc Lin, N.} 1993.
\newblock Cost analysis of logic programs.
\newblock {\em {ACM} Trans. Program. Lang. Syst.\/}~{\em 15,\/}~5, 826--875.

\bibitem[\protect\citeauthoryear{Debray, L{\'{o}}pez{-}Garc{\'{\i}}a,
  Hermenegildo, and Lin}{Debray
  et~al\mbox{.}}{1994}]{DBLP:conf/sas/DebrayGHL94}
{\sc Debray, S.~K.}, {\sc L{\'{o}}pez{-}Garc{\'{\i}}a, P.}, {\sc Hermenegildo,
  M.~V.}, {\sc and} {\sc Lin, N.} 1994.
\newblock Estimating the computational cost of logic programs.
\newblock In {\em Proc. {SAS} 1994}. LNCS, vol. 864. Springer, 255--265.

\bibitem[\protect\citeauthoryear{Flores{-}Montoya}{Flores{-}Montoya}{2017}]{DBLP:phd/dnb/Montoya17}
{\sc Flores{-}Montoya, A.} 2017.
\newblock Cost analysis of programs based on the refinement of cost relations.
\newblock Ph.D. thesis, Darmstadt University of Technology, Germany.

\bibitem[\protect\citeauthoryear{Flores{-}Montoya and
  H{\"{a}}hnle}{Flores{-}Montoya and
  H{\"{a}}hnle}{2014}]{DBLP:conf/aplas/Flores-MontoyaH14}
{\sc Flores{-}Montoya, A.} {\sc and} {\sc H{\"{a}}hnle, R.} 2014.
\newblock Resource analysis of complex programs with cost equations.
\newblock In {\em Proc. {APLAS} 2014}. LNCS, vol. 8858. Springer, 275--295.

\bibitem[\protect\citeauthoryear{Garcia, Laneve, and Lienhardt}{Garcia
  et~al\mbox{.}}{2015}]{DBLP:conf/ppdp/GarciaLL15}
{\sc Garcia, A.}, {\sc Laneve, C.}, {\sc and} {\sc Lienhardt, M.} 2015.
\newblock Static analysis of cloud elasticity.
\newblock In {\em Proc. {PPDP} 2015}. ACM, 125--136.

\bibitem[\protect\citeauthoryear{Giesl, Thiemann, Schneider-Kamp, and
  Falke}{Giesl et~al\mbox{.}}{2004}]{DBLP:conf/rta/GieslTSF04}
{\sc Giesl, J.}, {\sc Thiemann, R.}, {\sc Schneider-Kamp, P.}, {\sc and} {\sc
  Falke, S.} 2004.
\newblock Automated termination proofs with aprove.
\newblock In {\em Proc. {RTA} 2004}. LNCS, vol. 3091. Springer, Aachen,
  Germany, 210--220.

\bibitem[\protect\citeauthoryear{Grech, Georgiou, Pallister, Kerrison, Morse,
  and Eder}{Grech et~al\mbox{.}}{2015}]{DBLP:conf/scopes/GrechGPKME15}
{\sc Grech, N.}, {\sc Georgiou, K.}, {\sc Pallister, J.}, {\sc Kerrison, S.},
  {\sc Morse, J.}, {\sc and} {\sc Eder, K.} 2015.
\newblock Static analysis of energy consumption for {LLVM} {IR} programs.
\newblock In {\em Proc. {SCOPES} 2015}. ACM, 12--21.

\bibitem[\protect\citeauthoryear{Gulwani, Jain, and Koskinen}{Gulwani
  et~al\mbox{.}}{2009}]{DBLP:conf/pldi/GulwaniJK09}
{\sc Gulwani, S.}, {\sc Jain, S.}, {\sc and} {\sc Koskinen, E.} 2009.
\newblock Control-flow refinement and progress invariants for bound analysis.
\newblock In {\em Proc. of {PLDI} 2009}. ACM, 375--385.

\bibitem[\protect\citeauthoryear{Kafle, Gallagher, Gange, Schachte,
  S{\o}ndergaard, and Stuckey}{Kafle
  et~al\mbox{.}}{2018}]{DBLP:journals/tplp/KafleGGS18}
{\sc Kafle, B.}, {\sc Gallagher, J.~P.}, {\sc Gange, G.}, {\sc Schachte, P.},
  {\sc S{\o}ndergaard, H.}, {\sc and} {\sc Stuckey, P.~J.} 2018.
\newblock An iterative approach to precondition inference using constrained
  horn clauses.
\newblock {\em Theory Pract. Log. Program.\/}~{\em 18,\/}~3-4, 553--570.

\bibitem[\protect\citeauthoryear{Liqat, Georgiou, Kerrison,
  L{\'{o}}pez{-}Garc{\'{\i}}a, Gallagher, Hermenegildo, and Eder}{Liqat
  et~al\mbox{.}}{2015}]{DBLP:conf/fopara/LiqatGK0GHE15}
{\sc Liqat, U.}, {\sc Georgiou, K.}, {\sc Kerrison, S.}, {\sc
  L{\'{o}}pez{-}Garc{\'{\i}}a, P.}, {\sc Gallagher, J.~P.}, {\sc Hermenegildo,
  M.~V.}, {\sc and} {\sc Eder, K.} 2015.
\newblock Inferring parametric energy consumption functions at different
  software levels: {ISA} vs. {LLVM} {IR}.
\newblock In {\em Proc. {FOPARA} 2015, Selected Papers}. LNCS, vol. 9964.
  Springer, 81--100.

\bibitem[\protect\citeauthoryear{Navas, Mera, L{\'{o}}pez{-}Garc{\'{\i}}a, and
  Hermenegildo}{Navas et~al\mbox{.}}{2007}]{DBLP:conf/iclp/NavasMLH07}
{\sc Navas, J.~A.}, {\sc Mera, E.}, {\sc L{\'{o}}pez{-}Garc{\'{\i}}a, P.}, {\sc
  and} {\sc Hermenegildo, M.~V.} 2007.
\newblock User-definable resource bounds analysis for logic programs.
\newblock In {\em Proc. {ICLP} 2007}. LNCS, vol. 4670. Springer, 348--363.

\bibitem[\protect\citeauthoryear{Serrano, L{\'{o}}pez{-}Garc{\'{\i}}a, Bueno,
  and Hermenegildo}{Serrano
  et~al\mbox{.}}{2013}]{DBLP:journals/tplp/SerranoLBH13}
{\sc Serrano, A.}, {\sc L{\'{o}}pez{-}Garc{\'{\i}}a, P.}, {\sc Bueno, F.}, {\sc
  and} {\sc Hermenegildo, M.~V.} 2013.
\newblock Sized type analysis for logic programs.
\newblock {\em Theory Pract. Log. Program.\/}~{\em
  13,\/}~4-5-Online-Supplement.

\bibitem[\protect\citeauthoryear{Sharma, Dillig, Dillig, and Aiken}{Sharma
  et~al\mbox{.}}{2011}]{DBLP:conf/cav/SharmaDDA11}
{\sc Sharma, R.}, {\sc Dillig, I.}, {\sc Dillig, T.}, {\sc and} {\sc Aiken, A.}
  2011.
\newblock Simplifying loop invariant generation using splitter predicates.
\newblock In {\em Proc. of CAV 2011}. Springer, 703--719.

\bibitem[\protect\citeauthoryear{Sinn, Zuleger, and Veith}{Sinn
  et~al\mbox{.}}{2014}]{SZV14:CAV}
{\sc Sinn, M.}, {\sc Zuleger, F.}, {\sc and} {\sc Veith, H.} 2014.
\newblock A simple and scalable static analysis for bound analysis and
  amortized complexity analysis.
\newblock In {\em Proc. CAV 2014}. LNCS, vol. 8559. Springer, 745--761.

\bibitem[\protect\citeauthoryear{Spoto, Mesnard, and Payet}{Spoto
  et~al\mbox{.}}{2010}]{DBLP:journals/toplas/SpotoMP10}
{\sc Spoto, F.}, {\sc Mesnard, F.}, {\sc and} {\sc Payet, {\'{E}}.} 2010.
\newblock A termination analyzer for java bytecode based on path-length.
\newblock {\em {ACM} Trans. Program. Lang. Syst.\/}~{\em 32,\/}~3, 8:1--8:70.

\bibitem[\protect\citeauthoryear{Wegbreit}{Wegbreit}{1975}]{DBLP:journals/cacm/Wegbreit75}
{\sc Wegbreit, B.} 1975.
\newblock Mechanical {P}rogram {A}nalysis.
\newblock {\em Communications ACM\/}~{\em 18,\/}~9, 528--539.

\end{thebibliography}


\label{lastpage}
\end{document}